\documentclass[aps,superscriptaddress,amsmath,amssymb,floatfix,twocolumn,showpacs,amsfonts,longbibliography]{revtex4-1}

\usepackage{graphicx}
\usepackage{dcolumn}
\usepackage{bm}
\usepackage{amsmath}
\usepackage[colorlinks=true,bookmarks=false,citecolor=blue,linkcolor=blue,hyperfootnotes=true,urlcolor=blue]{hyperref}
\usepackage{color}
\usepackage{comment}
\usepackage{multirow}
\usepackage{makecell}
\usepackage{subscript}
\usepackage{xr-hyper}
\usepackage{hyperref}
\begin{document}

\title{Intertwined Magneto-Structural Criticality Driving Heterogeneous Collective-Pseudospin Dynamics in Sr$_2$IrO$_4$}

\author{A. Singh}
\email[Corresponding author: ]{arnabsingh21@gmail.com}
\affiliation{NSLS2, Brookhaven National Laboratory, Upton, NY 11973, USA}
\author{J. Kwon}
\affiliation{Department of Physics, Pohang University of Science and Technology, Pohang, South Korea}
\author{AM Milinda Abeykoon}
\affiliation{NSLS2, Brookhaven National Laboratory, Upton, NY 11973, USA}


\begin{abstract}
\vspace{0.1 cm}
Correlated quantum phenomena arising from spin-orbit-coupled degrees of freedom provide foundational paradigms in emerging quantum technologies and spin-based functionalities. Elucidating
the cooperative behaviors of these systems requires a precise probing of their intertwined
magnetic and structural configurations. Using complementary resonant X-ray diffraction and high-energy total scattering measurements on single crystals of $Sr_2IrO_4$, we observe that spontaneous structural distortions scale quadratically with the magnetic order parameter. Importantly, the lattice distortion inherits the characteristic \(2\mathrm{D}\ XY\)-like critical behavior of the underlying antiferromagnetic order. X-ray photon correlation spectroscopy (XPCS) reveals multiscale, temperature- and wave-vector-dependent magnetic dynamics characterized by compressed-exponential relaxation and signatures of critical slowing down near the magnetic transition.
These findings suggest the role of spin-orbit coupling and local structural distortions as the possible mechanism orchestrating such heterogeneous magnetic dynamics.

\end{abstract}

\maketitle

\section{Introduction}





 The convergence of topological electronic phases, magnetism, and electron correlations is evolving as a new frontier for the realization of unconventional states of condensed matter and next-generation technologies. Among strongly correlated electron systems, $5d$ transition-metal oxides (TMOs) provide a unique platform where the relativistic spin–orbit interaction competes directly with the electron correlation energy, stabilizing novel electronic and magnetic ground states without counterpart in conventional Mott insulators~\cite{annurev:/content/journals/10.1146/annurev-conmatphys-020911-125138, PhysRevLett.101.076402}. Although the spatially extended nature of $5d$ orbitals reduces the on-site Coulomb interaction ($U$) relative to $3d$ and $4d$ systems, it enhances hybridization and promotes strong orbital-dependent phenomena, including unconventional magnetism, orbital ordering, and topologically non-trivial phases such as Weyl semimetals or axion insulators~\cite{SqareLatticeIridates_bertinshaw2019square, SpinOrbitEntangled_minarro2024spin, SOI_arita2012ab}. Within this framework, the layered perovskite iridate $\text{Sr}_2\text{IrO}_4$ has emerged as a prototypical spin-orbit-assisted Mott insulator, where strong spin–orbit coupling stabilizes a $J_{\text{eff}}=1/2$ state featuring deeply entangled spin and orbital degrees of freedom~\cite{annurev:/content/journals/10.1146/annurev-conmatphys-020911-125138, Lu2020}. Consequently, $\text{Sr}_2\text{IrO}_4$ serves as a model system for exploring the emergence of exotic magnetism, electronic nematicity, and quantum phases driven by spin–orbit entanglement~\cite{Kim2024, Lu2020, PhysRevLett.106.136402, SIO_Growth_Heo2025, doi:10.1126/sciadv.adt5680}.

\begin{figure*}[t]
    \centering
    \includegraphics[width=0.85\textwidth]{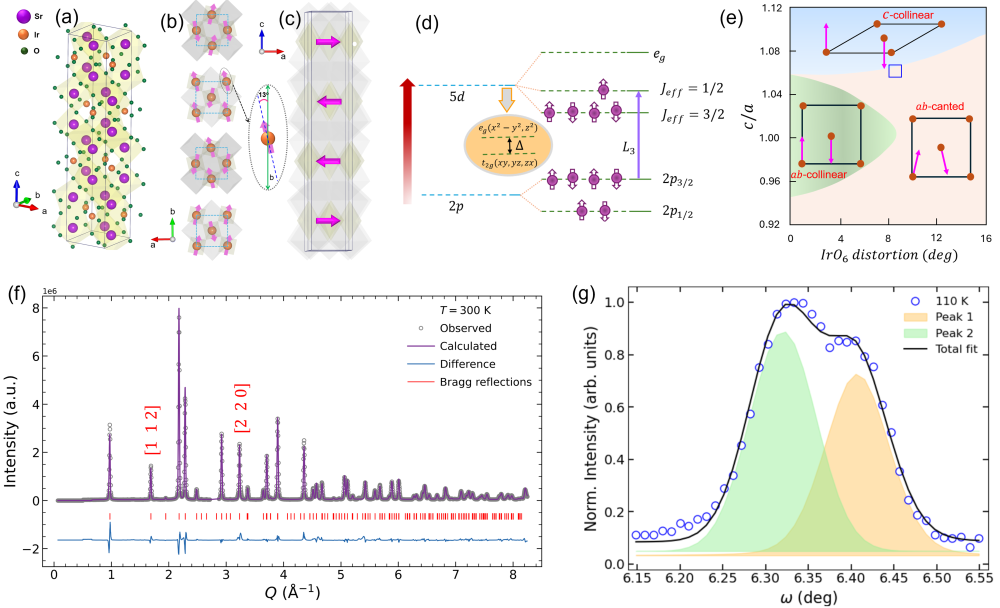}
     \caption{\textbf{Crystal structure, electronic profile, magnetic anisotropy, and diffraction profiles of \(\text{Sr}_2\text{Ir}\text{O}_4\).}
    \textbf{(a)} Crystal structure of $\text{Sr}_2\text{IrO}_4$. 
\textbf{(b)} In-plane canted antiferromagnetic (AFM) pseudospin arrangement. 
\textbf{(c)} Effective out-of-plane staggered AFM configuration. 
\textbf{(d)} Schematic of Ir $5d$ electronic level splitting into $J_{\text{eff}} = 1/2$ and $3/2$ states under crystal field and spin-orbit coupling, highlighting the $L_{3}$ resonant absorption transition. 
\textbf{(e)} Theoretical phase diagram illustrating the interplay among lattice parameters, octahedral rotations, and magnetic anisotropy~\cite{DFT_liu2015anisotropic}, with the blue box denoting the experimental parameter space of this work. \textbf{(f)} Room-temperature powder X-ray diffraction pattern confirming the $I4_{1}/acd$ space group. 
\textbf{(g)} Rocking curve of the $(1\,2\,0)$ AFM reflection revealing two components driven by distinct magnetic correlations.}
    \label{fig:Fig1}
\end{figure*}

In $\text{Sr}_2\text{IrO}_4$, comparable spin–orbit coupling, electron correlations, and crystal-field interactions stabilize a spin–orbit-entangled Mott insulating state~\cite{RXD_Sci_kim2009phase}.
Below the N\'eel temperature ($T_{\text{N}}\approx 240~K$), strong anisotropic exchange interactions—dictated by a staggered $\sim 13^\circ$ canting of the $J_{\mathrm{eff}}=1/2$ pseudospins along alternating clockwise and counterclockwise directions—establish a canted antiferromagnetic (AFM) ground state (Fig.~\ref{fig:Fig1}(b))~\cite{WeakFerro_cao1998weak, DFT_liu2015anisotropic, jackeli2009mott}. Structurally, these octahedral rotations lower the lattice symmetry, double the crystallographic unit cell ($I4_{1}/acd$ space group), and induce Brillouin zone folding (Figs.~\ref{fig:Fig1}(a--c))~\cite{FermiArcs, XRD_bhatti2015temperature, NeutronScattering_PhysRevB.87.144405, NeutronDiffraction_PhysRevB.87.140406}. Within the quasi-two-dimensional $\mathrm{IrO}_2$ planes, this structural distortion activates an antisymmetric Dzyaloshinskii-Moriya (DM) interaction that locks the staggered pseudospin canting within the $ab$-plane, breaking local time-reversal symmetry to generate an in-plane weak ferromagnetic moment. However, a staggered AFM stacking of adjacent layers (Fig.~\ref{fig:Fig1}(c)) completely cancels these components out-of-plane, restoring a compensated bulk magnetic structure with zero net spontaneous magnetization per unit cell~\cite{WeakFerro_cao1998weak, XRD_bhatti2015temperature}. This rigid locking among the spin, orbital, and lattice configurations renders the magnetic order highly susceptible to local structural modulations, while the underlying $\mathrm{IrO}_6$ distortions remain profoundly constrained by the thermodynamic evolution of the magnetic phase~\cite{2DXYFluctuations_PhysRevLett.84.3744, WeakFerro_cao1998weak, Pseudospin-lattice-coupling_PhysRevB.99.085125, BasalIn-planeAFM_PhysRevLett.110.117207, Lu2020}.

As illustrated in Fig.~\ref{fig:Fig1}(d), crystal-field splitting isolates the Ir $5d$ orbitals into $t_{2g}$ and $e_g$ manifolds, while large spin-orbit coupling (SOC) further splits the $t_{2g}$ states into a lower $J_{\mathrm{eff}}=3/2$ quartet and a higher $J_{\mathrm{eff}}=1/2$ doublet, establishing the electronic ground state of the system~\cite{EnergyBandchoi2024light, RXD_Sci_kim2009phase, SOI_arita2012ab}.
The resulting unpaired electron acts as a magnetic pseudospin, locking the magnetic moments directly to local structural variations (Fig.~\ref{fig:Fig1}(e)). 
Crucially, this intimate cross-coupling extends far beyond the static ground state, generating fundamentally intertwined spin and lattice fluctuations that dictate the multi-scale dynamical response of the system across its phase transitions~\cite{annurev:/content/journals/10.1146/annurev-conmatphys-020911-125138, SqareLatticeIridates_bertinshaw2019square, DualDynamics-Li_2019}. 
Pronounced quasi-2D critical fluctuations persist well into the paramagnetic (PM) regime ($T > T_{\text{N}}$), driven by strong XY anisotropy within the $J_{\text{eff}}=1/2$ manifold (Fig.~\ref{fig:Fig1}(e))~\cite{quasi2D_criticalfluc_fujiyama2012two, XY-anisotropy_PhysRevB.92.020406}. 
Although static pseudospin-lattice coupling is well documented~\cite{RXD_Sci_kim2009phase, SqareLatticeIridates_bertinshaw2019square, TheoreticalModel_PhysRevB.94.224420, MagnetoElasticCoupling_chikara2010giant}, its manifestation within equilibrium thermodynamics, highlighting how the resulting structural and magnetic criticality transform the underlying microscopic pseudospin fluctuation landscape to dictate the dynamics of mesoscopic domain boundaries in a spin-orbit Mott insulator, remains unexplored.

To address this question, we employ resonant X-ray diffraction, high-energy total X-ray scattering, and X-ray photon correlation spectroscopy (XPCS) to explore the magneto-structural criticality and spatio-temporal fluctuations. Below $T_{\mathrm{N}} \approx 240~\text{K}$, the system establishes long-range in-plane antiferromagnetic (AFM) order, while weak interlayer coupling drives a subsequent bulk 3D AFM order below $T_{\mathrm{3D}} \approx 225~\text{K}$. Our critical scaling analysis reveals that magnetically induced octahedral rotations and lattice distortions scale quadratically with the antiferromagnetic order parameter, exhibiting critical behavior near $T_N$ consistent with the 2D $XY$ universality class. XPCS measurements reveal heterogeneous and compressed nature of the underlying magnetic domain-wall dynamics. A slow relaxation channel ($\sim 10^{3}$~s) predominantly associated with 3D magnetic correlations, whereas a faster channel ($\sim 10^{1}$~s) is likely associated with correlations exhibiting 2D $XY$-like critical behavior. The compressed nature of the intermediate scattering function (ISF) provides further evidence of this coupled, collective pseudospin dynamics.

 \section{Methods}
\textbf{Growth:}
Single crystals were grown using ﬂux method. Stoichiometric amounts of IrO$_{2}$ and SrCO$_{3}$ powders were mixed with SrCl$_{2}$ flux. The powder-to-flux molar ratio was 1:7, based on previously reported synthesis methods~\cite{SIO_Growth_Heo2025}. 
The powder samples used for the total scattering measurements were obtained by grinding the as-grown single crystals.
\\

\textbf{Experimental Techniques:}

\textbf{Resonant X-ray diffraction (RXD) and X-ray photon correlation spectroscopy (XPCS)} measurements were performed at the Coherent Hard X-ray (CHX, 11-ID) beamline of the National Synchrotron Light Source II (NSLS-II), Brookhaven National Laboratory, using single-crystal of Sr$_2$IrO$_4$ near the Ir $L_3$ absorption edge to investigate the spatial magnetic correlations of the antiferromagnetic order. The sample was attached to the sample mount of a closed-cycle cryostat. The temperature-dependent RXD measurements were performed with an unfocused $\sim200~\mu\mathrm{m}$ beam over 100--300~K. 
The XPCS measurements were performed under resonant conditions using a $\sim50~\mu\mathrm{m}$ focused beam, and  a Eiger~1M area detector with 75  $\mu m$ pixel size, placed 150 cm downstream of the sample. At each measurement temperature, the sample was allowed to thermally equilibrate for 1~h after reaching the target temperature before XPCS data acquisition. Time resolved x-ray speckle patterns at different temperatures were captured for a total time of 30 min at a frame rate of 1 Hz.\\

\textbf{High-energy X-ray powder diffraction (XPD) and total-scattering measurements} were performed at the PDF beamline (28-ID-1) of the National Synchrotron Light Source II (NSLS-II), Brookhaven National Laboratory, using finely ground Sr$_2$IrO$_4$ powder (particle size $<20~\mu\mathrm{m}$) from the same sample batch used for the RXD and XPCS measurements. The powder sample was loaded into a Kapton tube and housed within a liquid helium cryostat for precise temperature regulation. Measurements were performed at 75~keV over the temperature range 15--375~K in steps of 5 K and the scattered x-rays was captured by a PerkinElmer area detector.  

\section{Background}
The Landau free energy per unit volume near the magnetic transition temperature ($T_C$) is expressed by \cite{landau2013course}: $\label{eqn:Landau Free Energy}
f = a(T - T_{\mathrm{C}})M^2
+ \frac{b}{2}M^4
- MB
+ \lambda \epsilon M^2
+ \frac{1}{2} C_{\mathrm{EL}}\epsilon^2 $,
where ${B}$ is the external magnetic field, ${M}$ is the sample magnetization, $\epsilon$ is the strain tensor, and $C_{EL}$ represents the elastic stiffness tensor. 
The fourth term ($\lambda_{} \epsilon_{} M^2$) describes the lowest-order magnetoelastic coupling between the strain and the magnetization.
The fifth term ($\frac{1}{2} C_{\text{EL}} \epsilon^2$)
accounts for the purely elastic energy contribution to the free energy. 

In the vicinity of the continuous magnetic transition ($T_{\mathrm{C}}$), the
equilibrium strain $\epsilon$ and magnetic order parameter $M$ are obtained
by minimizing the Landau free energy with respect to the corresponding
thermodynamic variables \cite{MagnetoLatticeStrain_PhysRevMaterials.5.064409}, and one obtains:
\begin{equation}
\label{eqn:magneto-elactic coupling}
    \epsilon = NM^2,
\end{equation}
\begin{equation}
\label{eq:order parameter}
M=A_0(T_{\mathrm{C}}-T)^{\beta},
\end{equation}

where $N$ defines the magnetostriction constant and $A_0$ is the scaling amplitude. 
See Sec.II in Supplementary Materials for the derivation of the equilibrium magnetization ($M$) and strain ($\epsilon$) from the Landau free energy expression. In $\text{Sr}_2\text{Ir}\text{O}_4$, the tight coupling between the $J_{\text{eff}} = 1/2$ magnetic moments and the tetragonal lattice symmetry makes this magnetostriction highly anisotropic~\cite{PhaseTransition_goldenfeld2018lectures, MagnetoElasticCoupling_chikara2010giant}. 

\begin{figure}[t]
    \centering    \includegraphics[width=\columnwidth]{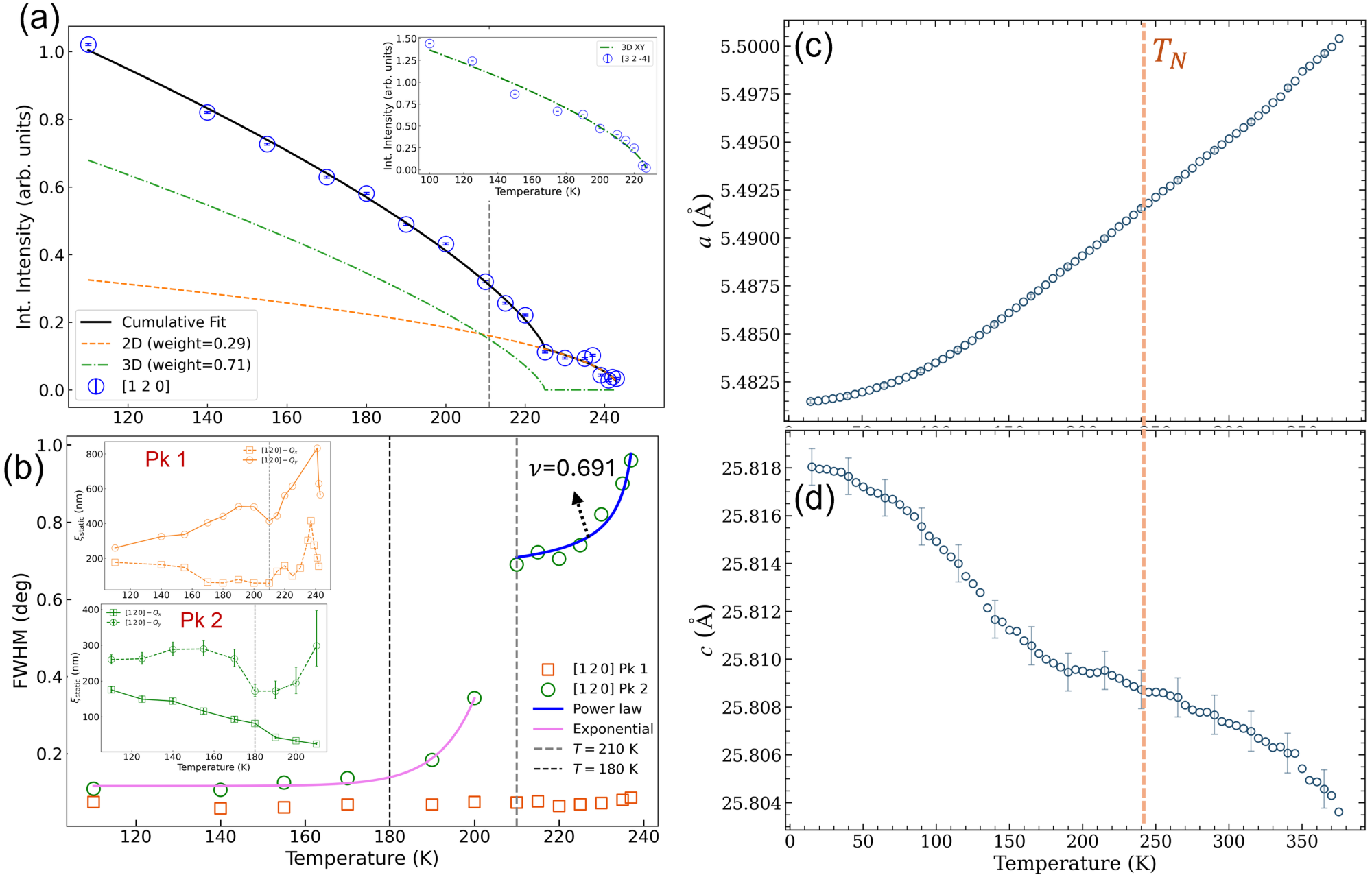}
     \caption{\textbf{Temperature dependent evolution of the AFM and structural order.}
    \textbf{(a)} Integrated area of the $(1\,2\,0)$ AFM rocking curve versus temperature, fitted to a composite power-law model (solid black curve) combining 2D-like (dashed orange) and 3D-like (dash-dotted green) magnetic correlations. Inset: Equivalent evolution and 3D critical fit (dotted green) for the $(3\,2\,\bar{4})$ AFM peak. 
\textbf{(b)} Rocking-scan FWHM temperature dependence with corresponding exponential and power-law fits. Power-law exponent value indicated by dashed arrow. Insets: Static correlation lengths ($\xi_{\mathrm{static}}$) along the detector-plane momentum-transfer directions $Q_x$ ($x-$direction) and $Q_y$ ($y-$direction) for Peak~1 (top) and Peak~2 (bottom) of the $(1\,2\,0)$ AFM reflection. Here, $Q_y$ lies in the scattering plane defined by the incident wave vector $\mathbf{k}_{\mathrm{in}}$ and the scattering vector $\mathbf{Q}_{(1\,2\,0)}$. The dashed grey and black lines in the top and bottom insets mark 210~K and 180~K, respectively.
Temperature evolution of the \textbf{(c)} $a$-axis and \textbf{(d)} $c$-axis lattice parameters.
}
    \label{fig:Fig2}
\end{figure}

\section{Results}
\subsection{Resonant X-ray Diffraction Analysis}
The $(1\,2\,0)$ magnetic Bragg reflection is predominantly sensitive to in-plane antiferromagnetic (AFM) ordering periodicity~\cite{NeutronDiffraction_PhysRevB.87.140406, TwinPeaks_[120]_PhysRevX.7.041018}. In contrast, the $(3\,2\,\bar{4})$ magnetic reflection is sensitive to the three-dimensional antiferromagnetic ordering periodicity, incorporating both in-plane magnetic correlations and interlayer magnetic coherence between adjacent $\mathrm{IrO}_2$ planes. The integrated magnetic Bragg scattering intensity ($I$) is proportional to the square of the staggered magnetization ($M$), which serves as the antiferromagnetic order parameter ($I\propto M^2$).

\subsubsection{Temperature dependence of the $(1\,2\,0)$ and $(3\,2\,\bar{4})$ AFM order}
 Figures~\ref{fig:Fig2}(a,b) show the temperature evolution of the integrated intensity and FWHM for the $(1\,2\,0)$ and $(3\,2\,\bar{4})$ antiferromagnetic (AFM) reflections. Above $T_{\mathrm{3D}} \approx 226$~K, the bulk 3D long-range order collapses, causing the $(3\,2\,\bar{4})$ reflection to vanish. In contrast, a robust component of the $(1\,2\,0)$ reflection persists up to $\sim 240$~K~\cite{quasi2D_criticalfluc_fujiyama2012two, jackeli2009mott}. As illustrated in Fig.~\ref{fig:Fig1}(g), the $(1\,2\,0)$ rocking curve comprises two distinct components: a low-angle peak (Pk2) that disappears above 225~K and the high-angle peak (Pk1) that survives to 240~K. The resulting temperature-dependent integrated intensities (background subtracted) are plotted over the 100--243~K range in Fig.~\ref{fig:Fig2}(a) (with the $(3\,2\,\bar{4})$ integrated intensity in the inset). To extract the effective dimensionality of magnetic interaction and magnetic anisotropy of these underlying magnetic correlations near the transition, the order parameter is modeled using the composite critical scaling relation of Eq.~\ref{eq:order parameter}:

\begin{equation}
\label{eqn:Eq. 1}
\begin{aligned}
I(T) = C + A\left(T_{\mathrm{N}}-T\right)^{2\beta_{1}} + B\left(T_{\mathrm{3D}}-T\right)^{2\beta_{2}}
\end{aligned}
\end{equation}
\\

Scaling analysis was performed over the 100--250 K range with $T_{\mathrm{N}}$ and $T_{\mathrm{3D}}$ as fitting parameters, along with a constant background. To track the temperature-dependent evolution of the peak profile, the integrated intensity was calculated directly from the raw normalized data where the background baseline corresponds to a constant value of \(I = 1.0\). Consequently, as the temperature approaches the highest measured temperature and the reflection completely suppresses, the peak profile nearly flattens into the baseline (see SM Fig. 4). The integrated intensity variation with temperature as plotted in Fig.~\ref{fig:Fig2}(a) is subtracted from the integrated intensity at the highest measured temperature for each of these magnetic reflections at all temperature points, thereby establishing a clear reference background for the nearly vanished order parameter at the transition temperature. The fitting exponents $\beta_{1}$ and $\beta_{2}$ quantify the critical behavior of the in-plane (2D) and  three dimensional magnetic correlations, respectively, while $A$ and $B$ denote their relative intensities. Fitting the $(1\,2\,0)$ integrated intensity profile (Fig.~\ref{fig:Fig2}(a)) via Eq.~\ref{eqn:Eq. 1} accurately decouples the quasi-2D-like and 3D-like magnetic correlations of the bulk system~\cite{XY-anisotropy_PhysRevB.92.020406}, yielding $T_{\mathrm{N}} \approx 240$~K and $T_{\mathrm{3D}} \approx 225$~K. The temperature-dependent $(3\,2\,\bar{4})$ integrated intensity (Fig.~\ref{fig:Fig2}(a)-inset) yields a critical exponent consistent with the 3D $XY$ universality class. The extracted critical exponents and transition temperatures are summarized in Table~\ref{tab:magnetic_critical_parameters}. 

The critical exponent $\beta_2$ extracted from this analysis falls within the standard limits of 3D $XY$ ($\beta \approx 0.345$), and 3D Heisenberg ($\beta \approx 0.365$) behaviors, essentially confirming 3D criticality~\cite{PhaseTransition_goldenfeld2018lectures, 3DIsing_3DHeisenberg_vale2019critical}. Remarkably, the $(1\,2\,0)$ component (orange dashed fit in Fig~\ref{fig:Fig2}(a)) that persists above $T_{\mathrm{3D}}$ is well described by an exponent of $\beta_1 \approx 0.25$. This value is close to the 2D $XY$ ($\beta \approx 0.26$) universality class~\cite{PhaseTransition_goldenfeld2018lectures, quasi2D_criticalfluc_fujiyama2012two, 2DXY_kosterlitz1974critical}. Our analysis identifies a  crossover temperature ($T_{\mathrm{2D-3D}}\approx210$~K), below which enhanced interlayer magnetic coupling favors predominantly 3D correlations~\cite{quasi2D_criticalfluc_fujiyama2012two, jackeli2009mott}. Near $T_{\mathrm{N}}\approx240$~K, the observed 2D $XY$-like behavior indicates that the effective dimensionality of magnetic interactions is predominantly two-dimensional ($d=2$), with pseudospins exhibiting easy-plane anisotropy ($XY$) within the $\mathrm{IrO}_2$ layers. Below $T_{\mathrm{3D}}\approx225$~K, the emergence of 3D $XY$-like behavior suggests that interlayer coupling extends the effective dimensionality of magnetic correlations across all three spatial directions ($d=3$), while preserving the easy-plane pseudospin anisotropy. The crossover near $T_{\mathrm{2D-3D}}\approx210$~K indicates the development of long range three-dimensional magnetic coherence throughout the bulk system.


\begin{table}[t]
\centering
\caption{Critical-scaling exponents and transition temperatures obtained from power-law fits to the integrated intensities of the $(1\,2\,0)$ and $(3\,2\,\bar{4})$ AFM reflections.}
\label{tab:magnetic_critical_parameters}

\renewcommand{\arraystretch}{1.8}
\setlength{\tabcolsep}{4pt}

\resizebox{\columnwidth}{!}{%
\begin{tabular}{c|ccccc}
\hline

\textbf{AFM Reflection} &
$\beta_1$ &
$T_N$ &
$\beta_2$ &
$T_{3D}$ &
$T_{2D-3D}$ \\

\hline

$(1\,2\,0)$ &
$0.25\pm0.06$ &
$244\pm0.8~\mathrm{K}$ &
$0.36\pm0.08$ &
$225\pm0.03~\mathrm{K}$ &
$210~\mathrm{K}$ \\

\hline

$(3\,2\,\bar{4})$ &
$-$ &
$-$ &
$0.334\pm0.007$ &
$227\pm0.03~\mathrm{K}$ &
$-$ \\

\hline

\end{tabular}%
}
\end{table}

To elucidate the nature of the critical disordering, we plot the temperature evolution of the FWHM for the quasi-2D (Pk1) and 3D (Pk2) effective magnetic interaction components of the $(1\,2\,0)$ reflection (Fig.~\ref{fig:Fig2}(b)) as obtained from the scaling analysis. The rocking-curve width is sensitive to magnetic domain misorientation and spatial disorder, potentially reflecting the influence of interlayer magnetic coupling on the antiferromagnetic domain structure. The quasi-2D component exhibits a sharp, nearly temperature-independent, FWHM with negligible broadening up to $T_{\mathrm{N}}$. In contrast, the 3D-associated component exhibits a pronounced change in its temperature dependence above $T_{\mathrm{2D-3D}}\approx210$~K (green open squares in Fig.~\ref{fig:Fig2}(b)). Below this crossover, the FWHM follows an exponential temperature dependence, tracking the systematic development of spatial order upon cooling. Above $\sim 210$~K, the FWHM increases sharply toward $T_{\mathrm{N}}$ via a power-law scaling that signifies scale-invariant critical disorder due to the collapse of the 3D interlayer staggered AFM stacking order~\cite{PhaseTransition_goldenfeld2018lectures, Stanley_RevModPhys.71.S358, 2DXYFluctuations_PhysRevLett.84.3744, quasi2D_criticalfluc_fujiyama2012two, DimensionalCrossover_PhysRevB.78.235118, Singh2019}. The insets of Fig.~\ref{fig:Fig2}(b) show the static correlation lengths, estimated as $\xi = 2\pi/\mathrm{FWHM}$ from the corresponding reciprocal-space peak profiles under specular conditions. After correcting for instrumental broadening using the direct-beam profile, the $Q_x$ and $Q_y$ correlation lengths of Peak~1 exhibit similar temperature dependence. In contrast, Peak~2 shows a pronounced reduction in the $Q_x$ correlation length above $\sim180$~K, coinciding with the onset of FWHM broadening (black dashed line in Fig.~\ref{fig:Fig2}(b)), consistent with a gradual weakening of inter-layer magnetic coupling.

The slight angular offset between the two overlapping profiles of the  $(1\,2\,0)$ reflection suggests a possible coexistence of distinct orientational textures within the domain matrix. This spatial arrangement may allow structurally tilted domains to independently develop different sensitivities to the predominantly two-dimensional in-plane magnetic interactions and three-dimensional interlayer-coupled magnetic correlations below the transition ($T_N$). We speculate that the persistence of 2D $XY$ domain behavior even below 210~K may be associated with local strain, local structural distortions, or interstitial defects that modify the interlayer exchange pathways by altering local bond lengths. This could locally weaken the effective interlayer coupling between spin–orbit-entangled \(J_{\mathrm{eff}}=1/2\) pseudospins, thereby limiting the development of three-dimensional magnetic correlations and allowing predominantly two-dimensional correlations to persist. A similar temperature-independent narrow FWHM component coexisting with a temperature-dependent broader component was previously reported by Vale \textit{et al.}~\cite{XY-anisotropy_PhysRevB.92.020406} and attributed to extrinsic structural defects. Our analysis further shows that the integrated intensity of the narrow component follows a scaling form consistent with 2D $XY$-like critical behavior. The progressive reduction in magnetic intensity with increasing temperature, accompanied by an approximately temperature-independent rocking-curve width from this peak, suggests enhanced thermal fluctuations of the spin orientations within magnetic domains, while the angular distribution of the scattering domains remains largely unchanged.

\begin{figure}[t]
    \centering
    \includegraphics[width=1.0\columnwidth]{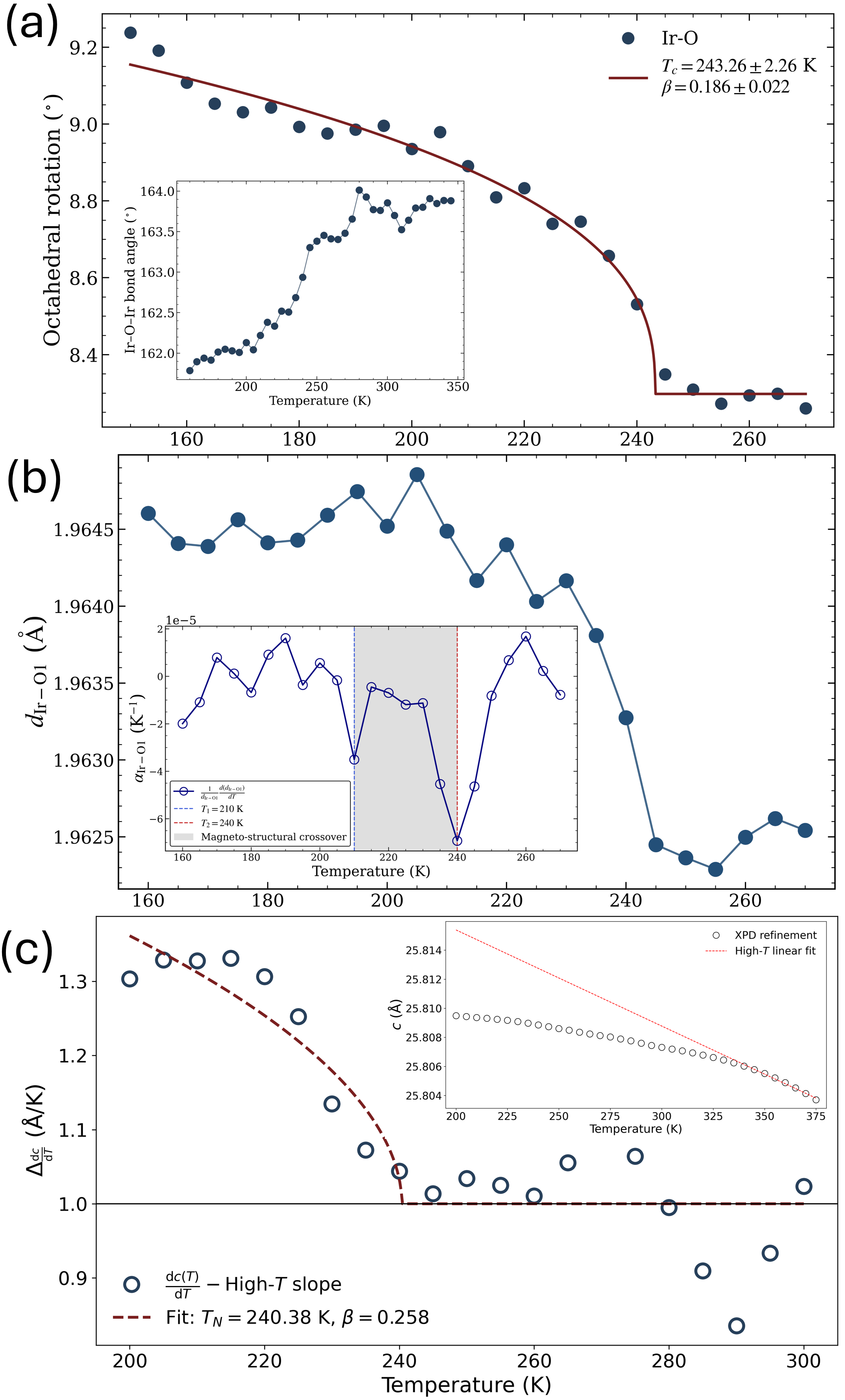}
    \caption{\textbf{Temperature evolution and critical fit of octahedral rotation, basal $Ir-O1$ bond-length and derivative, and $c$-axis lattice strain in $\mathrm{Sr_2IrO_4}$.}
    \textbf{(a)} Octahedral rotation as a function of temperature. The solid red line represents a power-law fit [Eq.~\ref{eq:order parameter}]. Inset shows the temperature dependence of the in-plane $\mathrm{Ir-O1-Ir}$ bond angle.
    \textbf{(b)} Basal \(\mathrm{Ir}\)–\(\mathrm{O1}\) bond length (\(d_{\mathrm{Ir-O1}}\)) as a function of temperature across the magnetic transitions. Inset shows the thermal expansion coefficient (\(\mathrm{\alpha}_{\mathrm{Ir-O1}}\)), revealing two distinct minimum at \(T_{\mathrm{3D}} \approx 210~\text{K}\) and \(T_{\mathrm{N}} \approx 240~\text{K}\) which bounds the magneto-structural crossover region (shaded area).
    \textbf{(c)} Evolution of the spontaneous magnetoelastic strain, $\Delta({\frac{dc}{dT}})$, that follows a critical power-law behavior (dashed red fitting curve). Inset shows the $c$-axis lattice parameter (open circles), compared with a high-temperature linear thermal-expansion fit (dashed red line)~\cite{PhysRevLett.111.096404}.}
    \label{fig:Fig3}
\end{figure}


\subsection{X-ray Powder Diffraction Analysis}
The temperature dependent lattice parameters ($a$ and $c$)
extracted by Rietveld refinement are shown in Figs.~\ref{fig:Fig2}(c, d), respectively. 
In the flux-grown crystals investigated here, both the in-plane ($a$) and out-of-plane ($c$) lattice parameters evolve continuously from $375$ down to $15~\mathrm{K}$. Details regarding the unit-cell parameters (\(c/a\) ratio and volume), apical bond length, and tetragonal distortions are provided in the SM (Fig.~5 in SM).

\subsubsection{Temperature dependence of the IrO$_6$ octahedral rotation and $Ir-O1$ basal bond length}
As shown in Figs.~\ref{fig:Fig2}(d) and \ref{fig:Fig3}(a,b), the thermal evolution of the $c$-lattice parameters, the basal bond length ($Ir-O1$) and the $IrO_6$ octahedral rotation capture a robust signature of magneto-structural coupling below $T_N$. The $c$-axis lattice parameter exhibits a distinct anomalous response within the 200--240~K range, which coincides with the onset of antiferromagnetic (AFM) order at $T_N$~\cite{XPD_g3cv-g5zj, RefinementXRD_PhysRevB.49.9198}.

Below $\sim 250$~K, the in-plane basal $\mathrm{Ir-O1-Ir}$ bond angle drops sharply untill 150~K, signaling enhanced staggered octahedral rotations (Fig.~\ref{fig:Fig3}(a)-inset). Concurrently, the basal bond length increases abruptly below $\sim 250$~K before plateauing below $\sim 200$~K (Fig.~\ref{fig:Fig3}(b)). Treating this bond length as a structural thermodynamic parameter, its first derivative behaves as a local thermal expansion coefficient ($\alpha_{\mathrm{local}} \equiv \mathrm{d}(d_{Ir-O1})/\mathrm{d}T$). Within the framework of the Ehrenfest relations, a continuous magnetic phase transition accompanied by appreciable spin--lattice coupling can produce an anomaly in the thermal expansion coefficient associated with the Ir--O1 bond length ($\alpha_{\mathrm{Ir-O1}}$). This anomaly may exhibit a temperature dependence correlated with the magnetic contribution to specific heat ($C_{\mathrm{mag}}$) and the temperature derivative of magnetic susceptibility ($d\chi/dT$), reflecting the coupling between magnetic ordering and lattice distortions~\cite{PhaseTransition_goldenfeld2018lectures, Stanley_RevModPhys.71.S358, MagnetoElasticCoupling_chikara2010giant, XRD_bhatti2015temperature}. 

The extracted \(\mathrm{\alpha}_{\mathrm{Ir-O1}}\) (see Fig.~\ref{fig:Fig3}(b)-inset) exhibit a sharp negative drop to its absolute minimum precisely at $T_{\mathrm{N}} = 240$~K. Upon cooling below $T_{\mathrm{N}}$ through the shaded magneto-structural crossover region, the derivative increases before showing a second pronounced dip around $T_{\mathrm{2D-3D}} = 210$~K. Crucially, this structural leveling directly matches the crossover to the 3D long-range antiferromagnetic interaction driven by the interlayer coupling.

\subsubsection{Modeling of the IrO$_6$ octahedral rotation and spontaneous $c$-axis strain across $T_N$}

The intrinsic geometric locking between the in-plane pseudospin-canting angle ($\phi$) and the structural octahedral rotation angle ($\theta_{\mathrm{oct}}$) establishes $\theta_{\mathrm{oct}}$ as a direct structural indicator for the magnetic configuration~\cite{Pseudospin-lattice-coupling_PhysRevB.99.085125, jackeli2009mott, DFT_PhysRevB.92.054428}. As shown in Fig.~\ref{fig:Fig3}(a), $\theta_{\mathrm{oct}}$ remains nearly constant upon cooling until it undergoes an abrupt, continuous expansion below $\sim 245$~K. 

To investigate the temperature-dependent evolution of the octahedral rotation below $\sim245$~K and its nature of coupling to the AFM order parameter, we perform a critical scaling analysis of the magnetically induced rotation-angle change using $\Delta\theta_{\mathrm{oct}}\propto(T_N-T)^{2\beta}$. The fit yields $T_N=243$~K and $\beta=0.186$, consistent with the critical exponent associated with the 2D $XYh_4$ universality class. This behavior supports a quadratic coupling between the octahedral distortion and the magnetic order parameter ($\Delta\theta_{\mathrm{oct}}\propto M^2$), as expected from minimizing the magnetoelastic free energy within Ginzburg--Landau theory~\cite{PhysRevB.79.184519_MagnetoStructuralCoupling, XY-anisotropy_PhysRevB.92.020406, LatticeParameters_Sharma_2022}. This analysis provides a quantitative characterization of the magnetostructural scaling in $\mathrm{Sr_2IrO_4}$. The 2D $XYh_4$ model is highly relevant here, as strong pseudospin-lattice coupling introduces a fourfold anisotropy breaking symmetry ($h_4$) term that locks spin orientations along specific crystallographic axes, lifting the continuous in-plane spin-rotational symmetry~\cite{XY-anisotropy_PhysRevB.92.020406, quasi2D_criticalfluc_fujiyama2012two, 2DXY_kosterlitz1974critical}. Moreover, the co-evolving octahedral rotation and tetragonal distortion (apical to  basal bond length ratio) places this system near the phase boundary separating the $ab$-canted and $c$-collinear magnetic regimes (blue box in Fig.~\ref{fig:Fig1}(e)).

\begin{table}[t]
\centering
\caption{Critical-scaling exponents and transition temperatures obtained from power-law fits to the structural distortions associated with octahedral rotations and $c$-lattice strain.}
\label{tab:structural_critical_parameters}

\renewcommand{\arraystretch}{1.8}
\setlength{\tabcolsep}{4pt}

\begin{tabular}{c|cc}
\hline

\textbf{Structural Distortion} &
$\beta_1$ &
$T_N$ \\

\hline

Octahedral Rotations &
$0.186\pm0.022$ &
$243.26\pm2.26~\mathrm{K}$ \\

\hline

$c$-Lattice Strain &
$0.258\pm0.11$ &
$240.38\pm1.87~\mathrm{K}$ \\

\hline

\end{tabular}

\end{table}

\begin{figure*}[t]
    \centering
    \includegraphics[width=1.0\textwidth]{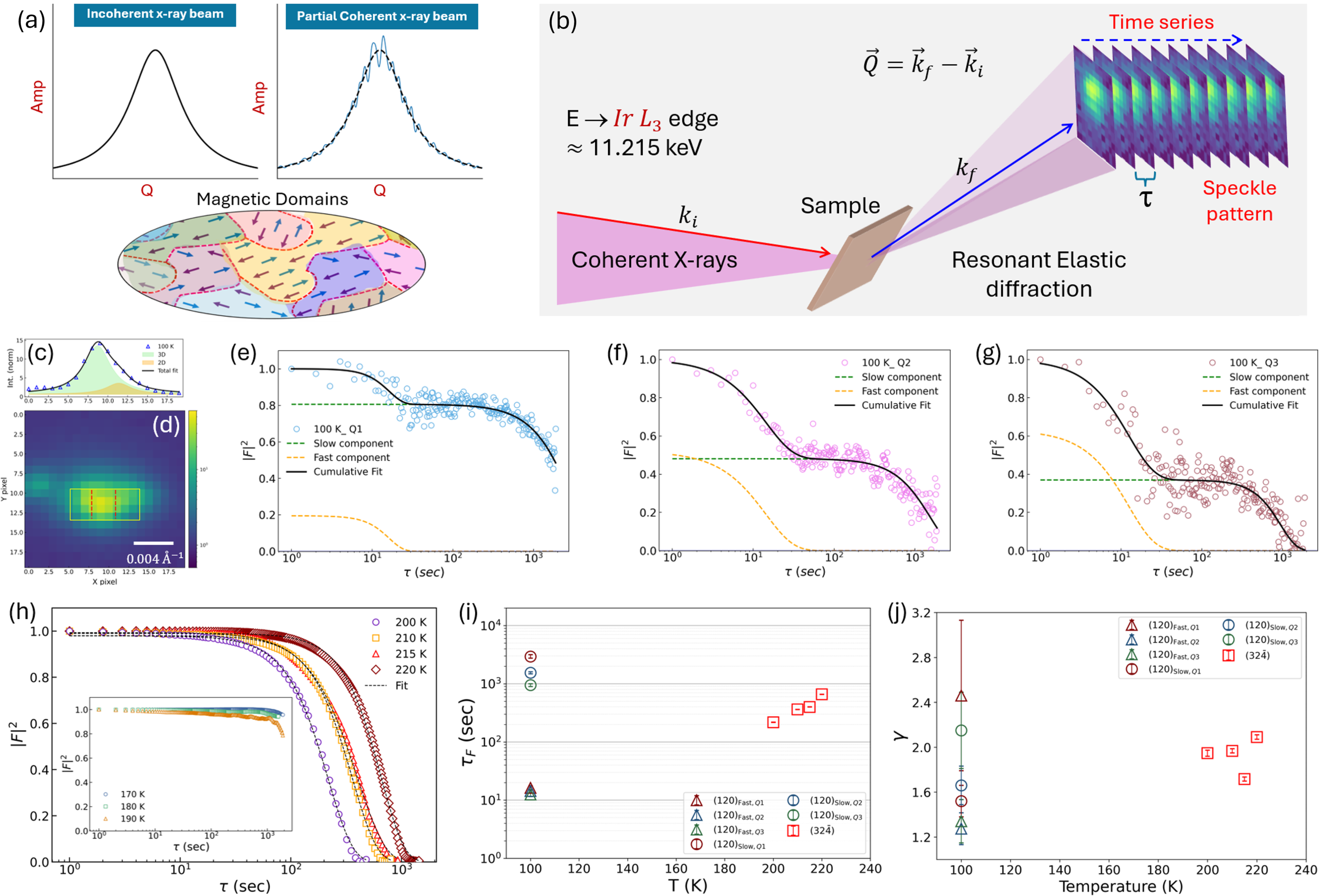}
     \caption{\textbf{Temperature-dependent X-ray Photon Correlation Spectroscopy (XPCS) of the antiferromagnetic Bragg reflection $\bf(1\,2\,0)$ and $\bf(3\,2\,\bar{4})$.} \textbf{(a)} Incoherent versus partially coherent scattering profiles over textured magnetic domains; coherent illumination yields fine interference sub-peaks due to spatial heterogeneity. 
\textbf{(b)} Resonant XPCS layout at the Ir $L_3$ edge showing a time series of magnetic speckle patterns recorded at intervals of $\tau$. The beam scatters from the sample with momentum transfer $\vec{Q} = \vec{k}_f - \vec{k}_i$. 
\textbf{(c, d)} Specular $(1\,2\,0)$ Bragg reflection profile and 2D scattering image at 100~K with ROI masks used for XPCS analysis. The shaded profiles (green for 3D and orange for quasi-2D magnetic correlations) denote the peak locations along the x-pixel array. 
\textbf{(e–g)} Intermediate scattering functions ($|F|^2$) versus delay time $\tau$ at 100~K across momentum regions ($Q_1$, $Q_2$, $Q_3$) marked by the yellow boundaries in panels (d). Data are modeled via using a multi-component decay combining slow dynamic (green dashed) and fast dynamic (orange dashed) components with the cumulative fit (black solid). 
\textbf{(h)} Compressed-exponential fits (dashed lines) to the $(3\,2\,\bar{4})$ AFM ISF profiles in the high-temperature (200--220~K) range and the low-temperature (170--190~K) regime shown in the inset. \textbf{(i, j)} Temperature dependence of the extracted characteristic time constants ($\tau_F$) and stretched exponents ($\gamma$) respectively from the distinct magnetic channels.}
    \label{fig:Fig4}
\end{figure*}

To further examine the evidence of magnetostructural coupling and its nature, we extracted the anomalous out-of-plane lattice strain ($\Delta c_{\mathrm{mag}}$) by subtracting the extrapolated high-temperature ($T>340$~K) nonmagnetic linear baseline from the measured $c$-axis parameter (Fig.~\ref{fig:Fig3}(c), inset)~\cite{PhysRevLett.111.096404}. As shown in Fig.~\ref{fig:Fig3}(c), $\Delta c_{\mathrm{mag}}$ develops near $T_{\mathrm{N}}$ and increases upon cooling, following a critical scaling form $\Delta c_{\mathrm{mag}}\propto(T_N-T)^{2\beta}$ with $\beta=0.258$. This behavior is consistent with a quadratic dependence on the AFM order parameter ($\Delta c_{\mathrm{mag}}\propto M^2$), providing further complementary evidence of magnetostructural coupling in $\mathrm{Sr_2IrO_4}$.
Mechanistically, the continuous contraction of the in-plane ($a$) lattice parameter brings adjacent in-plane Ir sites closer together, optimizing the $\mathrm{Ir-O1-Ir}$ basal bond geometry and enhancing $5d-2p$ orbital overlap~\cite{Hybridization_parschke2022evolution, ReviewIrridates_cao2018challenge, Pseudo-Jahn-Teller_PhysRevLett.122.057203}. 

 \subsection{Dynamics of the Magnetic order probed by X-ray Photon Correlation Spectroscopy}

As schematized in Fig.~\ref{fig:Fig4}(a), while conventional incoherent illumination yields a spatially averaged scattering profile, a partial coherent beam (available at several 3$^{rd}$ and 4$^{th}$ generation synchrotron X-ray sources) resolves the mesoscopic domain texture of the AFM order. The scattered wavefield forms a complex speckle pattern composed of fine interference sub-peaks, whose density scales with the number of distinct magnetic domains within the 3D coherence volume. These sharp features originate from phase heterogeneity (local variations in magnetization) and spatial heterogeneity (local magnetic lattice disorder) distributed across the Ir-spin lattice network.

In thermodynamic equilibrium, thermal fluctuations drive the temporal evolution of these domain wall  boundaries. By tracking this variance via X-ray Photon Correlation Spectroscopy (XPCS), the intensity-intensity autocorrelation function ($g_2$) captures the stochastically evolving speckle pattern. The correlation characteristic times ($\tau_{F}$) describe the collective lifetime of the domain configuration.
From these steady-state time-series ($10^0-10^3$~s), the intermediate scattering function, $\lvert F(Q,\tau)\rvert^2$, was extracted from $g_2(Q,\tau)$ via the relation~\cite{g2_PhysRevE.76.010401, g2_PhysRevLett.110.217201}:
$\label{eqn:g2}
 g2(Q,\tau) = \frac{\langle I(Q,t^\prime)I(Q,t^\prime +\tau)\rangle_{t^\prime}}{\langle I(Q,t^\prime)\rangle^{2}_{t^\prime}}=\beta\lvert F(Q,\tau)\rvert^{2} + g_{\infty}
$, where $I(Q,t^\prime)$ is the intensity of a speckle at wave vector Q at a time $t^\prime$, $\tau$ is the time delay, and angular brackets represent averages over the sub-scripted variable. ${F(Q,\tau)}$ is the intermediate scattering function, $\beta$ is the amplitude of the speckle contrast, and the baseline ($g_{\infty}$) for $\tau \to \infty$. For intensity autocorrelation functions $g_2(\tau)$ that do not reach a well-defined long-time plateau within the experimental time frame, the asymptotic baseline $g_{\infty}$ is taken to be unity for simplicity, in accordance to the expected normalization of the intensity autocorrelation function in the limit of complete decorrelation. The above equation auto-correlates speckle intensities at wavevector ${Q}$ with different lag times ($\tau$) and provides insight into the length-scale dependent dynamic behavior of the AFM domains.\\
Figures~\ref{fig:Fig4} (e-h) show the intermediate scattering function, \(|F(Q,\tau)|^{2}\), obtained from the \((1\,2\,0)\) and \((3\,2\,\bar{4})\) AFM Bragg reflections over a range of temperatures below $T_N$. At 100~K and across the Q-regions, the dynamics exhibit two distinct characteristic timescales for the \((1\,2\,0)\) peak, indicating the coexistence of more than one dynamical processes. We therefore model the measured intermediate scattering function (${F(Q,\tau)}$) using a double-stretched-exponential form ~\cite{compositeExponentialFUnction_shpyrko2007direct}:
$|F({Q},\tau)|^2
=
a\exp\left[-\left(\frac{\tau}{\tau_{\mathrm{F_1}}}\right)^{\gamma_1}\right]
+
(1-a)\exp\left[-\left(\frac{\tau}{\tau_{\mathrm{F_2}}}\right)^{\gamma_2}\right] $,
where $\tau_{F_1}$ and $\tau_{F_2}$ represent the characteristic fast and slow fluctuation timescales with corresponding exponents $\gamma_1$ and $\gamma_2$ having amplitudes $a$ and $1-a$ receptively. The value of the exponent ($\gamma$) was found to be $> 1$ for all the cases, indicating a compressed exponential function. A compressed exponent, \(\gamma>1\), is generally associated with non-diffusive, spatially correlated relaxation and is often observed in systems where relaxation proceeds through cooperative rearrangements, internal stress relaxation, or strain-mediated dynamics. In contrast, \(\gamma=1\) corresponds to a simple exponential relaxation, whereas \(\gamma<1\) generally indicates a broad distribution of relaxation times ~\cite{Jamming_PhysRevLett.121.177601}. 

\subsubsection{Fluctuation dynamics of the $(1\,2\,0)$ AFM order}
Figure~\ref{fig:Fig4}(e-g) presents the ISF, $|F(\mathbf{q},\tau)|^{2}$, plotted against delay time $\tau$ across distinct momentum transfer regions ($Q_1$, $Q_2$, $Q_3$) of the $(1\,2\,0)$ antiferromagnetic (AFM) reflection at 100~K. To decouple the dynamics of the overlapping 3D-like magnetic order and 2D-like in-plane magnetic correlations, diffracted intensity profiles from three continuous regions of interest (ROIs; Figs.~\ref{fig:Fig4}(d)) were fitted simultaneously using a composite pseudo-Voigt function (Figs.~\ref{fig:Fig4}(c)). As static measurements at the specular condition (see SM Fig.~S6) revealed that the magnetic scattering component due to the 3D effective magnetic interactions ((Pk~2) dominates over 2D-like behavior (Pk~1) at 100~K, hence the low-$Q$ region ($Q_1$) is primarily sensitive to 3D correlations, the high-$Q$ region ($Q_3$) captures more 2D-like correlations, and the center ($Q_2$) represents a mixed regime in Fig.~\ref{fig:Fig4}(c,d).

Modeling the corresponding $|F(\mathbf{q},\tau)|^{2}$ curves using a double-stretched-exponential function reveals a systematic redistribution of the relative contributions of the two relaxation channels, with the fast component becoming increasingly dominant over the slow component as the wave vector shifts from $Q_1$ to $Q_3$ (see Figs.~\ref{fig:Fig4}(e--g). This evolution is consistent with a progressively greater contribution from quasi-2D magnetic fluctuations relative to 3D magnetic correlations, as illustrated in Figs.~\ref{fig:Fig4}(c,d). The results indicate that slow domain-wall dynamics (\(\sim 10^{3}\)~s) has a higher contribution at the reciprocal space (\(Q\)-region) dominated by 3D magnetic correlations (green shaded profile). However, the distinct and significantly faster dynamical channel (\(\sim 10^{1}\)~s) carries greater weight in the \(Q\)-region where quasi-2D in-plane correlations dominate (orange shaded profile). This interpretation is consistent with previous time-resolved resonant inelastic X-ray scattering (RIXS) studies on \(\mathrm{Sr}_2\mathrm{IrO}_4\), which revealed faster recovery timescales for (\(2\mathrm{D}\)) in-plane correlations than (\(3\mathrm{D}\)) long-range magnetic order~\cite{UltrafastRIXS_dean2016ultrafast}. However, because the fluctuation timescales in our XPCS measurements are inherently convoluted by the magneto-structural coupling at the probed lengthscales, assigning these dual dynamical channels strictly to the intrinsic magnetic correlation dimensionalities remains highly suggestive rather than definitive. At $200$~K and above, the overall scattering intensity decreases substantially as the system approaches the crossover associated with the progressive loss of 3D magnetic coherence, thereby reducing the reliability of the extracted fitting parameters. Across the \(Q\)-regions of interest, the contribution associated with 3D magnetic correlations progressively diminishes from left to right, whereas predominantly quasi-2D in-plane correlations persist and become increasingly dominant in the overall magnetic scattering signal above $200$~K (see Supplementary Fig.~6 for the corresponding profiles at 200~K and 205~K). 
The extracted compressed exponents $\gamma$ at 100~K exhibit values between $\sim$1.3 and $\sim$2.4. The observation of exponents $\gamma > 1$ indicates that the measured magnetic fluctuations cannot be described by independent, localized diffusive processes, but instead signify correlated collective dynamics~\cite{g2_PhysRevLett.110.217201, Jamming_PhysRevLett.121.177601}. A compressed exponential shape ($\gamma > 1$) represents collective relaxation or pinning-induced jamming behavior, where local magnetic stress fields or domain boundary constraints yield long-range spatial inhomogeneities ~\cite{Jamming_PhysRevLett.121.177601, CompressedExponenet1.5_PhysRevLett.93.228302}. This jamming behavior ($\gamma > 1$) is often interpreted as an ultraslow, ballistic-like motion of magnetic domain boundaries or spin textures under the influence of elastic relaxation from heterogeneous local internal stresses~\cite{g2_PhysRevLett.110.217201, compositeExponentialFUnction_shpyrko2007direct}. A universal signature of such collective relaxation has been widely documented in soft and hard condensed matter systems ~\cite{Jamming_PhysRevLett.121.177601,SoftMatterXPCS_fluerasu2010dynamics,CDW_PhysRevLett.117.167001, g2_PhysRevLett.110.217201, compositeExponentialFUnction_shpyrko2007direct, StretchedExponent_frey2025liquid}.

\subsubsection{Fluctuation dynamics of the \((3\,2\,\bar{4})\) AFM order}
The temperature-dependent dynamics associated with 3D magnetic order, probed within the high-intensity ROI (85th--98th percentile), exhibit nearly static $|F(\mathbf{q},\tau)|^{2}$ at 170--190~K, with relaxation times exceeding the experimental window ($\sim 10^{3}$~s), followed by pronounced decorrelation upon warming toward $T_{\mathrm{2D-3D}}$ (Fig.~\ref{fig:Fig4}(h)).  
To model this transition into a single dominant dynamic channel, the $|F(\mathbf{q},\tau)|^{2}$ curves were fitted using a single-component decay function (with $a=1$ in the double-stretched-exponential form). The correlation curves are characterized by a compressed exponential with $\gamma \approx 1.7\text{--}2.1$, signaling strongly jammed, spatially correlated, and non-diffusive dynamics~\cite{CompressedExponent1.8_PhysRevLett.122.107802, compositeExponentialFUnction_shpyrko2007direct, StretchedExponent_frey2025liquid}. The ISF associated with the $(3\,2\,\bar{4})$ magnetic reflection exhibits complete decorrelation within the experimental observation window at temperatures between 190 and 200~K, near the onset of the dimensional crossover regime ($T_{\mathrm{2D-3D}}$). Detailed values for extracted characteristic timescales and compressed exponents from the fittings are summarized in Figs.~\ref{fig:Fig4}(i) and \ref{fig:Fig4}(j) respectively. Above 200~K, $\tau_F$ increases systematically toward $T_{\mathrm{3D}}\approx225$~K, indicating a progressive slowing of magnetic fluctuations consistent with critical slowing down near the three-dimensional magnetic ordering transition. A common feature of all measured $|F(\mathbf{q},\tau)|^2$ curves is their compressed-exponential relaxation, quantitatively characterized by a relaxation exponent $\gamma>1$. The characteristic relaxation times ($\tau_F$) and compressed-exponential exponents ($\gamma$) extracted from the $(3\,2\,\bar{4})$ reflection are comparable in magnitude to those associated with the predominantly 3D-like correlations of the $(1\,2\,0)$ reflection, despite their different scattering geometries and sensitivities to magnetic correlations. While the former is sensitive to the magnetic order involving both in-plane and out-of-plane components, the latter primarily probes in-plane magnetic order, suggesting that the observed relaxation dynamics may share some common underlying mechanism owing to the inter-layer coupling.


\section{Discussions}
The nature of how magnetic fluctuations evolves with temperature is dictated by several key underlying mechanisms: (i) the competition between intralayer and interlayer magnetic exchange interactions in $\mathrm{Sr}_{2}\mathrm{IrO}_{4}$, (ii) the magneto-structural coupling between canted $J_{\mathrm{eff}} = 1/2$ pseudospins and $\mathrm{IrO}_{6}$ octahedral rotations, and (iii) critical thermal fluctuations upon approaching the N\'{e}el temperature $T_{\mathrm{N}}$. Below $T_{\mathrm{3D}}$, interlayer magnetic correlations establish three-dimensional antiferromagnetic order, coupling the in-plane pseudospin configurations of adjacent $\mathrm{IrO}_{2}$ layers through the staggered anti-ferromagnetic stacking along the $c$-axis. However, spatial variations in the effective interlayer coupling may produce distinct domain-wall dynamics, with regions less constrained by interlayer correlations exhibiting faster, predominantly 2D-like fluctuations, while strongly correlated regions display slower, 3D-like dynamics. Such dynamical heterogeneity is consistent with the observed distinct characteristic timescales associated with the 2D $XY$-like and 3D $XY$-like components of the $(1\,2\,0)$ reflection at $100$~K. The development of long-range interlayer coherence supports a crossover toward 3D critical behavior near $T_{\mathrm{3D}}$, below which 3D antiferromagnetic order is established. Above this temperature, thermal fluctuations suppress interlayer couplings, driving a crossover toward predominantly 2D in-plane magnetic interactions characterized by absence of magnetic interactions between adjacent $\mathrm{IrO}_{2}$ layers~\cite{quasi2D_criticalfluc_fujiyama2012two, BasalIn-planeAFM_PhysRevLett.110.117207}. The power-law critical exponent $\nu\approx0.69$, extracted from the temperature-dependent FWHM of the fluctuating magnetic domains near the critical temperature $T_N$(Fig.~\ref{fig:Fig2}(b)), lies between the theoretical 3D $XY$ ($\nu\approx0.67$) and 3D Heisenberg ($\nu\approx0.71$) values. This intermediate value reflects the evolving effective magnetic anisotropy and effective dimensionality of the magnetic interactions within an unit cell.

\subsubsection{Theoretical Framework}

To understand how coupled degrees of freedom influence the associated dynamics, we employ a theoretical framework based on linear response theory and the generalized fluctuation-dissipation theorem (FDT)~\cite{Kubo_1966}. We define an interaction Hamiltonian density $H_{\text{int}} = -\int \mathbf{g} \tilde{S}(\mathbf{r}) u(\mathbf{r}) \, d\mathbf{r}$, coupling ($\mathbf{g}$) the pseudospin density $\tilde{S}(\mathbf{r}, t)$ to a localized structural coordinate $u(\mathbf{r}, t)$.
By utilizing linear response theory to expand this cross-susceptibility to the lowest order in $\mathbf{g}$ as the direct product of the bare intrinsic spin ($\chi_{\tilde{S}\tilde{S}}^{(0)}$) and lattice ($\chi_{uu}^{(0)}$) response functions~\cite{FDT_coleman_linearresponse_2022}, we can isolate the dissipative component governing the cross-dynamics:
\begin{equation}
\begin{aligned}
\left\langle \delta\tilde{S}(\mathbf{q},\omega) \delta u(-\mathbf{q},-\omega) \right\rangle = & \frac{2\hbar \mathbf{g}}{1-e^{-\hbar\omega/(k_{\text{B}}T)}} \\
& \times \operatorname{Im}\Big[ \chi_{\tilde{S}\tilde{S}}^{(0)}(\mathbf{q},\omega) \cdot \chi_{uu}^{(0)}(\mathbf{q},\omega) \Big].
\end{aligned}
\label{eq:CrossFluctuations}
\end{equation}

For the full derivation see Sec. III of Supplementary Materials. Within this perturbative framework, Eq.~\eqref{eq:CrossFluctuations} demonstrates how a finite pseudospin-lattice coupling (${\bf{g}} \neq 0$) generates correlated dynamical fluctuations whose amplitude and frequency dependencies are governed by the intrinsic response functions of the coupled subsystems~\cite{FDT_coleman_linearresponse_2022, Kubo_1966}. This formalism provides a rigorous theoretical foundation for interpreting the dynamics resolved via steady-state XPCS measurements, mapping the relative weight of competing structural and magnetic modes to their temporal evolution via characteristic timescales (${\tau_F}$). Furthermore, the coupling parameter $\mathbf{g}$ provides a microscopic mechanism for collective relaxation processes that may contribute to the observed compressed-exponential behavior ($\gamma>1$). In the uncoupled limit ($\bf{g} = 0$), the subsystems fluctuate independently via classical thermodynamics, mathematically restricting the decay profile to the classical Brownian/Debye limit ($\gamma = 1$)~\cite{Kubo_1966}, whereas independent parallel relaxation pathways yield a stretched-exponential line shape ($\gamma < 1$)~\cite{palmer1984model}. Conversely, when $\bf{g} > 0$, localized microscopic spin reorientation act as elastic dipoles that generate long-range 
elastic strain fields, creating a jammed, collective network. Across all measured dynamical channels and antiferromagnetic reflections, the average value of the stretched exponent is $\langle\gamma\rangle\approx1.8$, indicating predominantly compressed-exponential dynamics. According to the canonical Cipelletti-Bouchaud model for stressed networks, the collective, ballistic-like relaxation and dissipation of internal structural stresses across the heterogeneous domain network mathematically force the system into a compressed exponential profile (\(\gamma > 1\)) with the characteristic signature of $\gamma \approx 1.5$ for strongly cooperative dynamics~\cite{cipelletti2000universal, cipelletti2003glassy, bouchaud2001anomalous, falcao2002slow, ElShawish_2018, compositeExponentialFUnction_shpyrko2007direct, CompressedExponenet1.5_PhysRevLett.93.228302}.

\subsubsection{Nature and Origin of the Dynamic Behavior}
For the $(1\,2\,0)$ reflection, the separation of the underlying magnetic domains allows the distinct speckles associated with the quasi-2D and 3D like magnetic interactions. Consequently, even when these distinct speckles overlap at the specular condition, they can partition into separate dynamical channels within the intermediate scattering function if the intrinsic fluctuations timescale are distinct between them. Conversely, for the $(3\,2\,\bar{4})$ reflection, the 3D correlation results in only one dynamical channel in the intermediate scattering function.
This multi-component and correlated dynamic landscape closely aligns with time-resolved resonant X-ray scattering (tr-REXS/tr-RIXS), magneto-optical, and ultrafast reflectivity measurements, which resolve a fast electronic recovery alongside a highly bottlenecked spin-lattice equilibration channel near the transition~\cite{DualDynamics-Li_2019, UltrafastRIXS_dean2016ultrafast, MOKE_PhysRevX.9.021020}.

\section{Conclusions}
We demonstrate through static and time-resolved measurements that local octahedral distortions are structurally locked to the canting of pseudospins, dynamically imprinting critical magnetic fluctuations onto the lattice framework. Mediated by the continuous corner-sharing network of the \(\mathrm{IrO}_{6}\) cages, these fluctuations manifest as a highly collective, coupled response~\cite{XPD_g3cv-g5zj}.
The intermediate scattering functions exhibit predominantly compressed-exponential relaxation and coexisting distinct timescales, revealing heterogeneous magnetic dynamics indicative of cooperative spin--lattice relaxation dynamics. The observed dynamics shows slow-jammed fluctuations possible originating from discrete domain-wall rearrangements within the coupled pseudospin-octahedral network, rather than the spatially homogeneous, continuous motion characteristic of independent fluctuations. These findings suggest that spatial and temporal heterogeneities are characteristic of slow, coupled magnetostructural dynamics in this spin-orbit-coupled system. The interplay of in-plane pseudospin exchange, weak interlayer coupling, and local structural distortions provides a framework for understanding its heterogeneous magnetic dynamics. More broadly, our complementary experimental approach enables the investigation of correlated fluctuations across extended spatial and temporal scales, offering insights into the dynamical behavior of strongly correlated quantum materials.

\section*{Acknowledgments}
This research used resources of the National Synchrotron Light Source II, a U.S. Department of Energy (DOE) Office of Science User Facility operated for the DOE Office of Science by Brookhaven National Laboratory under Contract No. DE-SC0012704.
We thank A. Fluerasu and X. M. Chen for their technical assistance and beamline support during the experiments at the CHX beamline of NSLS-II. The authors acknowledges Masafumi Fukuto for his review of the manuscript and valuable comments and suggestions. The authors declare no competing interests. 


\bibliographystyle{apsrev4-1}
\bibliography{main}

@article{Kubo_1966,
  author = {Kubo, R.},
  title = {The fluctuation-dissipation theorem},
  journal = {Reports on Progress in Physics},
  year = {1966},
  volume = {29},
  pages = {255--284},
  doi = {10.1088/0034-4885/29/1/306}
}

@article{PhysRevLett.101.076402,
  title = {Novel ${J}_{\mathrm{eff}}=1/2$ Mott State Induced by Relativistic Spin-Orbit Coupling in ${\mathrm{Sr}}_{2}{\mathrm{IrO}}_{4}$},
  author = {Kim, B. J. and Jin, Hosub and Moon, S. J. and Kim, J.-Y. and Park, B.-G. and Leem, C. S. and Yu, Jaejun and Noh, T. W. and Kim, C. and Oh, S.-J. and Park, J.-H. and Durairaj, V. and Cao, G. and Rotenberg, E.},
  journal = {Phys. Rev. Lett.},
  volume = {101},
  issue = {7},
  pages = {076402},
  numpages = {4},
  year = {2008},
  month = {Aug},
  publisher = {American Physical Society},
  doi = {10.1103/PhysRevLett.101.076402},
  url = {https://link.aps.org/doi/10.1103/PhysRevLett.101.076402}
}

@article{SIO_Growth_Heo2025,
  author   = {Heo, Hyeokjun and An, Jeongha and Kwon, Junyoung and Kim, Kwangrae and Son, Youngoh and Kim, B. J. and Jang, Joonho},
  title    = {Interplay of canted antiferromagnetism and nematic order in Mott insulating $\text{Sr}_2\text{Ir}_{1-x}\text{Rh}_x\text{O}_4$},
  journal  = {Communications Physics},
  year     = {2025},
  volume   = {8},
  number   = {1},
  pages    = {470},
  month    = {11},
  url      = {https://www.nature.com/articles/s42005-025-02372-y},
  issn     = {2399-3650}
}

@article{annurev:/content/journals/10.1146/annurev-conmatphys-020911-125138,
   author = "Witczak-Krempa, William and Chen, Gang and Kim, Yong Baek and Balents, Leon",
   title = "Correlated Quantum Phenomena in the Strong Spin-Orbit Regime", 
   journal= "Annual Review of Condensed Matter Physics",
   year = "2014",
   volume = "5",
   number = "Volume 5, 2014",
   pages = "57-82",
   doi = "https://doi.org/10.1146/annurev-conmatphys-020911-125138",
   url = "https://www.annualreviews.org/content/journals/10.1146/annurev-conmatphys-020911-125138",
   publisher = "Annual Reviews",
   issn = "1947-5462",
   type = "Journal Article",
  }

@article{SpinOrbitEntangled_minarro2024spin,
  title={Spin-orbit entanglement driven by the Jahn-Teller effect},
  author={Mi{\~n}arro, Alejandro S and Villa, Mario and Casals, Blai and Plana-Ruiz, Sergi and S{\'a}nchez, Florencio and G{\'a}zquez, Jaume and Herranz, Gervasi},
  journal={Nature communications},
  volume={15},
  number={1},
  pages={8694},
  year={2024},
  publisher={Nature Publishing Group UK London},
  url = {https://www.nature.com/articles/s41467-024-52848-8#citeas}
}

@article{SOI_arita2012ab,
  title={Ab initio Studies on the Interplay between Spin-Orbit Interaction<? format?> and Coulomb Correlation in Sr 2 IrO 4 and Ba 2 IrO 4},
  author={Arita, R and Kune{\v{s}}, Jan and Kozhevnikov, AV and Eguiluz, AG and Imada, M},
  journal={Physical review letters},
  volume={108},
  number={8},
  pages={086403},
  year={2012},
  publisher={APS},
  url = {https://journals.aps.org/prl/abstract/10.1103/PhysRevLett.108.086403}
}

@article{Kim2024,
  author  = {Kim, Hoon and Kim, Jin-Kwang and Kwon, Junyoung and Kim, Jimin and Kim, Hyun-Woo J. and Ha, Seunghyeok and Kim, Kwangrae and Lee, Wonjun and Kim, Jonghwan and Cho, Gil Young and Heo, Hyeokjun and Jang, Joonho and Sahle, C. J. and Longo, A. and Strempfer, J. and Fabbris, G. and Choi, Y. and Haskel, D. and Kim, Jungho and Kim, J.-W. and Kim, B. J.},
  title   = {Quantum spin nematic phase in a square-lattice iridate},
  journal = {Nature},
  year    = {2024},
  volume  = {625},
  number  = {7994},
  pages   = {264--269},
  doi     = {10.1038/s41586-023-06829-4},
  url     = {https://doi.org/10.1038/s41586-023-06829-4},
  issn    = {1476-4687}
}

@article{jackeli2009mott,
  title={Mott Insulators in the Strong Spin-Orbit Coupling Limit:<? format?> From Heisenberg to a Quantum Compass and Kitaev Models},
  author={Jackeli, George and Khaliullin, Giniyat},
  journal={Physical review letters},
  volume={102},
  number={1},
  pages={017205},
  year={2009},
  publisher={APS},
  url = {https://journals.aps.org/prl/abstract/10.1103/PhysRevLett.102.017205}
}

@article{Lu2020,
  author  = {Lu, Chengliang and Liu, Jun-Ming},
  title   = {The {$J_{\mathrm{eff}} = 1/2$} Antiferromagnet {Sr$_2$IrO$_4$}: A Golden Avenue toward New Physics and Functions},
  journal = {Advanced Materials},
  year    = {2020},
  volume  = {32},
  number  = {27},
  pages   = {1904508},
  doi     = {10.1002/adma.201904508},
  url     = {https://doi.org/10.1002/adma.201904508}
}

@article{quasi2D_criticalfluc_fujiyama2012two,
  title={Two-dimensional Heisenberg behavior of J eff= 1/2 isospins in the paramagnetic state of the spin-orbital Mott insulator Sr 2 IrO 4},
  author={Fujiyama, S and Ohsumi, H and Komesu, T and Matsuno, J and Kim, BJ and Takata, M and Arima, T and Takagi, H},
  journal={Physical review letters},
  volume={108},
  number={24},
  pages={247212},
  year={2012},
  publisher={APS},
  url = {https://journals.aps.org/prl/abstract/10.1103/PhysRevLett.108.247212}
}

@article{XY-anisotropy_PhysRevB.92.020406,
  title = {Importance of $XY$ anisotropy in ${\mathrm{Sr}}_{2}{\mathrm{IrO}}_{4}$ revealed by magnetic critical scattering experiments},
  author = {Vale, J. G. and Boseggia, S. and Walker, H. C. and Springell, R. and Feng, Z. and Hunter, E. C. and Perry, R. S. and Prabhakaran, D. and Boothroyd, A. T. and Collins, S. P. and R\o{}nnow, H. M. and McMorrow, D. F.},
  journal = {Phys. Rev. B},
  volume = {92},
  issue = {2},
  pages = {020406(R)},
  numpages = {5},
  year = {2015},
  month = {Jul},
  publisher = {American Physical Society},
  doi = {10.1103/PhysRevB.92.020406},
  url = {https://link.aps.org/doi/10.1103/PhysRevB.92.020406}
}

@article{RXD_Sci_kim2009phase,
  title={Phase-sensitive observation of a spin-orbital Mott state in Sr2IrO4},
  author={Kim, BJ and Ohsumi, H and Komesu, T and Sakai, S and Morita, T and Takagi, H and Arima, T-h},
  journal={Science},
  volume={323},
  number={5919},
  pages={1329--1332},
  year={2009},
  publisher={American Association for the Advancement of Science},
  url = {https://www.science.org/doi/full/10.1126/science.1167106}
}

@article{DFT_liu2015anisotropic,
  title={Anisotropic magnetic couplings and structure-driven canted to collinear transitions in Sr 2 IrO 4 by magnetically constrained noncollinear DFT},
  author={Liu, Peitao and Khmelevskyi, Sergii and Kim, Bongjae and Marsman, Martijn and Li, Dianzhong and Chen, Xing-Qiu and Sarma, DD and Kresse, Georg and Franchini, Cesare},
  journal={Physical Review B},
  volume={92},
  number={5},
  pages={054428},
  year={2015},
  publisher={APS},
  url = {https://journals.aps.org/prb/abstract/10.1103/PhysRevB.92.054428}
}

@article{XRD_bhatti2015temperature,
  title={Temperature evolution of magnetic and transport behavior in 5 d Mott insulator Sr2IrO4: significance of magneto-structural coupling},
  author={Bhatti, Imtiaz Noor and Rawat, R and Banerjee, A and Pramanik, AK},
  journal={Journal of physics: Condensed matter},
  volume={27},
  number={1},
  pages={016005},
  year={2015},
  publisher={IOP Publishing},
  url = {https://iopscience.iop.org/article/10.1088/0953-8984/27/1/016005/meta}
}

@article{PhysRevLett.106.136402,
  title = {Twisted Hubbard Model for ${\mathrm{Sr}}_{2}{\mathrm{IrO}}_{4}$: Magnetism and Possible High Temperature Superconductivity},
  author = {Wang, Fa and Senthil, T.},
  journal = {Phys. Rev. Lett.},
  volume = {106},
  issue = {13},
  pages = {136402},
  numpages = {4},
  year = {2011},
  month = {Mar},
  publisher = {American Physical Society},
  doi = {10.1103/PhysRevLett.106.136402},
  url = {https://link.aps.org/doi/10.1103/PhysRevLett.106.136402}
}

@article{Pseudospin-lattice-coupling_PhysRevB.99.085125,
  title = {Pseudospin-lattice coupling in the spin-orbit Mott insulator ${\mathrm{Sr}}_{2}{\mathrm{IrO}}_{4}$},
  author = {Porras, J. and Bertinshaw, J. and Liu, H. and Khaliullin, G. and Sung, N. H. and Kim, J.-W. and Francoual, S. and Steffens, P. and Deng, G. and Sala, M. Moretti and Efimenko, A. and Said, A. and Casa, D. and Huang, X. and Gog, T. and Kim, J. and Keimer, B. and Kim, B. J.},
  journal = {Phys. Rev. B},
  volume = {99},
  issue = {8},
  pages = {085125},
  numpages = {11},
  year = {2019},
  month = {Feb},
  publisher = {American Physical Society},
  doi = {10.1103/PhysRevB.99.085125},
  url = {https://link.aps.org/doi/10.1103/PhysRevB.99.085125}
}

@article{SqareLatticeIridates_bertinshaw2019square,
  author   = {Bertinshaw, Joel and Kim, Y. K. and Khaliullin, Giniyat and Kim, B. J.},
  title    = {Square Lattice Iridates},
  journal  = {Annual Review of Condensed Matter Physics},
  volume   = {10},
  number   = {1},
  pages    = {315--336},
  year     = {2019},
  doi      = {10.1146/annurev-conmatphys-031218-013113},
  url      = {https://doi.org/10.1146/annurev-conmatphys-031218-013113}
}

@article{TheoreticalModel_PhysRevB.94.224420,
  title = {Model analysis of magnetic susceptibility of ${\text{Sr}}_{2}{\text{IrO}}_{4}$: A two-dimensional ${J}_{\mathrm{eff}}=\frac{1}{2}$ Heisenberg system with competing interlayer couplings},
  author = {Takayama, Tomohiro and Matsumoto, Akiyo and Jackeli, George and Takagi, Hidenori},
  journal = {Phys. Rev. B},
  volume = {94},
  issue = {22},
  pages = {224420},
  numpages = {7},
  year = {2016},
  month = {Dec},
  publisher = {American Physical Society},
  doi = {10.1103/PhysRevB.94.224420},
  url = {https://link.aps.org/doi/10.1103/PhysRevB.94.224420}
}

@article{FermiArcs,
  author  = {Y. K. Kim and O. Krupin and J. D. Denlinger and A. Bostwick and E. Rotenberg and Q. Zhao and J. F. Mitchell and J. W. Allen and B. J. Kim},
  title   = {Fermi arcs in a doped pseudospin-1/2 Heisenberg antiferromagnet},
  journal = {Science},
  volume  = {345},
  number  = {6193},
  pages   = {187--190},
  year    = {2014},
  doi     = {10.1126/science.1251151},
  url     = {https://www.science.org/doi/abs/10.1126/science.1251151}
}

@article{WeakFerro_cao1998weak,
  title={Weak ferromagnetism, metal-to-nonmetal transition, and negative differential resistivity in single-crystal Sr 2 IrO 4},
  author={Cao, G and Bolivar, J and McCall, S and Crow, JE and Guertin, RP},
  journal={Physical Review B},
  volume={57},
  number={18},
  pages={R11039},
  year={1998},
  publisher={APS},
  url = {https://journals.aps.org/prb/abstract/10.1103/PhysRevB.57.R11039}
}

@article{EnergyBandchoi2024light,
  title={Light-induced insulator-metal transition in Sr$_2$IrO$_4$ reveals the nature of the insulating ground state},
  author={Choi, D. and Yue, C. and Azoury, D. and others},
  journal={Proc. Natl. Acad. Sci. U.S.A.},
  volume={121},
  number={29},
  pages={e2323013121},
  year={2024},
  doi={10.1073/pnas.2323013121}
}

@article{Pseudo-Jahn-Teller_PhysRevLett.122.057203,
  title = {Pseudo-Jahn-Teller Effect and Magnetoelastic Coupling in Spin-Orbit Mott Insulators},
  author = {Liu, Huimei and Khaliullin, Giniyat},
  journal = {Phys. Rev. Lett.},
  volume = {122},
  issue = {5},
  pages = {057203},
  numpages = {6},
  year = {2019},
  month = {Feb},
  publisher = {American Physical Society},
  doi = {10.1103/PhysRevLett.122.057203},
  url = {https://link.aps.org/doi/10.1103/PhysRevLett.122.057203}
}

@article{DualDynamics-Li_2019,
doi = {10.1088/1361-648X/ab123d},
url = {https://doi.org/10.1088/1361-648X/ab123d},
year = {2019},
month = {apr},
publisher = {IOP Publishing},
volume = {31},
number = {25},
pages = {255801},
author = {Li, Jiemin and Wang, Ruitang and Guo, Haizhong and Zhu, Yi and Cao, Yue and Liu, Jian and Ding, Hong and Wen, Haidan and Liu, X},
title = {Recovery of photoexcited magnetic ordering in Sr2IrO4},
journal = {Journal of Physics: Condensed Matter}
}

@book{PhaseTransition_goldenfeld2018lectures,
  title={Lectures on phase transitions and the renormalization group},
  author={Goldenfeld, Nigel},
  year={2018},
  publisher={CRC Press},
  url = {https://www.routledge.com/Lectures-On-Phase-Transitions-And-The-Renormalization-Group/Goldenfeld/p/book/9780429493492}
}

@article{2DXY_kosterlitz1974critical,
  title={The critical properties of the two-dimensional xy model},
  author={Kosterlitz, J Michael},
  journal={Journal of Physics C: Solid State Physics},
  volume={7},
  number={6},
  pages={1046--1060},
  year={1974},
  url = {https://iopscience.iop.org/article/10.1088/0022-3719/7/6/005/meta?casa_token=-EaVl9hL-4oAAAAA:stuldP0haDnfAKPanOBYMiqxRmGGfljPkRmRGcP8XG-LRmKMd8ydpLmsq29jrqcC0AimScf6feHYKwjFUdZhUZGHs-5URw}
}

@article{3DIsing_3DHeisenberg_vale2019critical,
  title={Critical fluctuations in the spin--orbit Mott insulator Sr3Ir2O7},
  author={Vale, JG and Boseggia, S and Walker, HC and Springell, RS and Hunter, EC and Perry, RS and Collins, SP and McMorrow, DF},
  journal={Journal of Physics: Condensed Matter},
  volume={31},
  number={18},
  pages={185803},
  year={2019},
  publisher={IOP Publishing},
  url = {https://iopscience.iop.org/article/10.1088/1361-648X/ab0471/meta}
}

@article{2DXYFluctuations_PhysRevLett.84.3744,
  title = {Universal Fluctuations in Correlated Systems},
  author = {Bramwell, S. T. and Christensen, K. and Fortin, J.-Y. and Holdsworth, P. C. W. and Jensen, H. J. and Lise, S. and L{\'o}pez, J. M. and Nicodemi, M. and Pinton, J.-F. and Sellitto, M.},
  journal = {Phys. Rev. Lett.},
  volume = {84},
  issue = {17},
  pages = {3744--3747},
  numpages = {0},
  year = {2000},
  month = {Apr},
  publisher = {American Physical Society},
  doi = {10.1103/PhysRevLett.84.3744},
  url = {https://link.aps.org/doi/10.1103/PhysRevLett.84.3744}
}

@article{DimensionalCrossover_PhysRevB.78.235118,
  title = {Dimensional crossover in quantum critical metallic magnets},
  author = {Garst, Markus and Fritz, Lars and Rosch, Achim and Vojta, Matthias},
  journal = {Phys. Rev. B},
  volume = {78},
  issue = {23},
  pages = {235118},
  numpages = {18},
  year = {2008},
  month = {Dec},
  publisher = {American Physical Society},
  doi = {10.1103/PhysRevB.78.235118},
  url = {https://link.aps.org/doi/10.1103/PhysRevB.78.235118}
}

@article{Singh2019,
  author    = {Singh, A. and Lee, J. C. T. and Avila, K. E. and Chen, Y. and Montoya, S. A. and Fullerton, E. E. and Fischer, P. and Dahmen, K. A. and Kevan, S. D. and Sanyal, M. K. and Roy, S.},
  title     = {Scaling of domain cascades in stripe and skyrmion phases},
  journal   = {Nature Communications},
  year      = {2019},
  volume    = {10},
  number    = {1},
  pages     = {1988},
  month     = {apr},
  doi       = {10.1038/s41467-019-09934-z},
  url       = {https://doi.org/10.1038/s41467-019-09934-z}
}

@article{Stanley_RevModPhys.71.S358,
  title = {Scaling, universality, and renormalization: Three pillars of modern critical phenomena},
  author = {Stanley, H. Eugene},
  journal = {Rev. Mod. Phys.},
  volume = {71},
  issue = {2},
  pages = {S358--S366},
  numpages = {0},
  year = {1999},
  month = {Mar},
  publisher = {American Physical Society},
  doi = {10.1103/RevModPhys.71.S358},
  url = {https://link.aps.org/doi/10.1103/RevModPhys.71.S358}
}

@article{DFT_PhysRevB.92.054428,
  title = {Anisotropic magnetic couplings and structure-driven canted to collinear transitions in ${\mathrm{Sr}}_{2}{\mathrm{IrO}}_{4}$ by magnetically constrained noncollinear DFT},
  author = {Liu, Peitao and Khmelevskyi, Sergii and Kim, Bongjae and Marsman, Martijn and Li, Dianzhong and Chen, Xing-Qiu and Sarma, D. D. and Kresse, Georg and Franchini, Cesare},
  journal = {Phys. Rev. B},
  volume = {92},
  issue = {5},
  pages = {054428},
  numpages = {6},
  year = {2015},
  month = {Aug},
  publisher = {American Physical Society},
  doi = {10.1103/PhysRevB.92.054428},
  url = {https://link.aps.org/doi/10.1103/PhysRevB.92.054428}
}

@book{landau2013course,
  title={Course of Theoretical Physics},
  volume={5},
  note={Statistical Physics, Part 1 (3rd Edition)},
  author={Landau, Lev Davidovich and Lifshitz, Evgenii Mikhailovich},
  year={2013},
  publisher={Elsevier / Pergamon},
  isbn={978-1-4831-0337-2},
  url={https://shop.elsevier.com/books/course-of-theoretical-physics/landau/978-0-08-023039-9}
}

@article{MagnetoElasticCoupling_chikara2010giant,
  title={Giant magnetoelectric effect in the {J_{\text{eff}}} = 1/2 {Mott} insulator {{Sr}_2{\text{IrO}}_4}},
  author={Chikara, S. and Korneta, O. and Crummett, W. P. and DeLong, L. E. and Schlottmann, P. and Cao, G.},
  journal={Journal of Applied Physics},
  volume={107},
  number={9},
  pages={09D910},
  year={2010},
  doi={10.1063/1.3362912},
  url={https://doi.org},
  publisher={AIP Publishing}
}

@article{MagnetoLatticeStrain_PhysRevMaterials.5.064409,
  title = {Magnetostriction of $\mathrm{Al}{\mathrm{Fe}}_{2}{\mathrm{B}}_{2}$ in high magnetic fields},
  author = {Sharma, S. and Kovalev, A. E. and Rebar, D. J. and Mann, D. and Yannello, V. and Shatruk, M. and Suslov, A. V. and Smith, J. H. and Siegrist, T.},
  journal = {Phys. Rev. Mater.},
  volume = {5},
  issue = {6},
  pages = {064409},
  numpages = {8},
  year = {2021},
  month = {Jun},
  publisher = {American Physical Society},
  doi = {10.1103/PhysRevMaterials.5.064409},
  url = {https://link.aps.org/doi/10.1103/PhysRevMaterials.5.064409}
}

@article{RefinementXRD_PhysRevB.49.9198,
  title = {Structural and magnetic studies of ${\mathrm{Sr}}_{2}$${\mathrm{IrO}}_{4}$},
  author = {Crawford, M. K. and Subramanian, M. A. and Harlow, R. L. and Fernandez-Baca, J. A. and Wang, Z. R. and Johnston, D. C.},
  journal = {Phys. Rev. B},
  volume = {49},
  issue = {13},
  pages = {9198--9201},
  numpages = {0},
  year = {1994},
  month = {Apr},
  publisher = {American Physical Society},
  doi = {10.1103/PhysRevB.49.9198},
  url = {https://link.aps.org/doi/10.1103/PhysRevB.49.9198}
}

@article{TwinPeaks_[120]_PhysRevX.7.041018,
  title = {Correlation Effects and Hidden Spin-Orbit Entangled Electronic Order in Parent and Electron-Doped Iridates ${\mathrm{Sr}}_{2}{\mathrm{IrO}}_{4}$},
  author = {Zhou, Sen and Jiang, Kun and Chen, Hua and Wang, Ziqiang},
  journal = {Phys. Rev. X},
  volume = {7},
  issue = {4},
  pages = {041018},
  numpages = {9},
  year = {2017},
  month = {Oct},
  publisher = {American Physical Society},
  doi = {10.1103/PhysRevX.7.041018},
  url = {https://link.aps.org/doi/10.1103/PhysRevX.7.041018}
}

@article{NeutronScattering_PhysRevB.87.144405,
  title = {Neutron scattering study of correlated phase behavior in Sr${}_{2}$IrO${}_{4}$},
  author = {Dhital, Chetan and Hogan, Tom and Yamani, Z. and de la Cruz, Clarina and Chen, Xiang and Khadka, Sovit and Ren, Zhensong and Wilson, Stephen D.},
  journal = {Phys. Rev. B},
  volume = {87},
  issue = {14},
  pages = {144405},
  numpages = {5},
  year = {2013},
  month = {Apr},
  publisher = {American Physical Society},
  doi = {10.1103/PhysRevB.87.144405},
  url = {https://link.aps.org/doi/10.1103/PhysRevB.87.144405}
}

@article{NeutronDiffraction_PhysRevB.87.140406,
  title = {Magnetic and crystal structures of ${\mathrm{Sr}}_{2}{\mathrm{IrO}}_{4}$: A neutron diffraction study},
  author = {Ye, Feng and Chi, Songxue and Chakoumakos, Bryan C. and Fernandez-Baca, Jaime A. and Qi, Tongfei and Cao, G.},
  journal = {Phys. Rev. B},
  volume = {87},
  issue = {14},
  pages = {140406(R)},
  numpages = {6},
  year = {2013},
  month = {Apr},
  publisher = {American Physical Society},
  doi = {10.1103/PhysRevB.87.140406},
  url = {https://link.aps.org/doi/10.1103/PhysRevB.87.140406}
}

@article{g2_PhysRevE.76.010401,
  title = {Slow dynamics and aging in colloidal gels studied by x-ray photon correlation spectroscopy},
  author = {Fluerasu, Andrei and Moussa{\"{\i}}d, Abdellatif and Madsen, Anders and Schofield, Andrew},
  journal = {Phys. Rev. E},
  volume = {76},
  issue = {1},
  pages = {010401(R)},
  numpages = {4},
  year = {2007},
  month = {Jul},
  publisher = {American Physical Society},
  doi = {10.1103/PhysRevE.76.010401},
  url = {https://link.aps.org/doi/10.1103/PhysRevE.76.010401}
}

@article{g2_PhysRevLett.110.217201,
  title = {Jamming Behavior of Domains in a Spiral Antiferromagnetic System},
  author = {Chen, S.-W. and Guo, H. and Seu, K. A. and Dumesnil, K. and Roy, S. and Sinha, S. K.},
  journal = {Phys. Rev. Lett.},
  volume = {110},
  issue = {21},
  pages = {217201},
  numpages = {5},
  year = {2013},
  month = {May},
  publisher = {American Physical Society},
  doi = {10.1103/PhysRevLett.110.217201},
  url = {https://link.aps.org/doi/10.1103/PhysRevLett.110.217201}
}

@article{compositeExponentialFUnction_shpyrko2007direct,
  title={Direct measurement of antiferromagnetic domain fluctuations},
  author={Shpyrko, O. G. and Isaacs, E. D. and Logan, J. M. and Feng, Yejun and Aeppli, G. and Jaramillo, Rafael and Kim, H. C. and Rosenbaum, T. F. and Zschack, Paul and Sprung, Michael and Narayanan, S. and Sandy, A. R.},
  journal={Nature},
  volume={447},
  number={7140},
  pages={68--71},
  year={2007},
  doi={10.1038/nature05776},
  url={https://www.nature.com/articles/nature05776#citeas},
  publisher={Nature Publishing Group}
}

@article{Jamming_PhysRevLett.121.177601,
  title = {Orbital Domain Dynamics in Magnetite below the Verwey Transition},
  author = {Kukreja, Roopali and Hua, Nelson and Ruby, Joshua and Barbour, Andi and Hu, Wen and Mazzoli, Claudio and Wilkins, Stuart and Fullerton, Eric E. and Shpyrko, Oleg G.},
  journal = {Phys. Rev. Lett.},
  volume = {121},
  issue = {17},
  pages = {177601},
  numpages = {6},
  year = {2018},
  month = {Oct},
  publisher = {American Physical Society},
  doi = {10.1103/PhysRevLett.121.177601},
  url = {https://link.aps.org/doi/10.1103/PhysRevLett.121.177601}
}

@article{CompressedExponenet1.5_PhysRevLett.93.228302,
  title = {Evolution of Particle-Scale Dynamics in an Aging Clay Suspension},
  author = {Bandyopadhyay, R. and Liang, D. and Yardimci, H. and Sessoms, D. A. and Borthwick, M. A. and Mochrie, S. G. J. and Harden, J. L. and Leheny, R. L.},
  journal = {Phys. Rev. Lett.},
  volume = {93},
  issue = {22},
  pages = {228302},
  numpages = {4},
  year = {2004},
  month = {Nov},
  publisher = {American Physical Society},
  doi = {10.1103/PhysRevLett.93.228302},
  url = {https://link.aps.org/doi/10.1103/PhysRevLett.93.228302}
}

@article{StretchedExponent_frey2025liquid,
  title={Liquid-like versus stress-driven dynamics in a metallic glass former observed by temperature scanning {X}-ray photon correlation spectroscopy},
  author={Frey, Maximilian and Neuber, Nico and Riegler, Sascha Sebastian and Cornet, Antoine and Chushkin, Yuriy and Zontone, Federico and Ruschel, Lucas Matthias and Gallino, Isabella and Busch, Ralf and Ruta, Beatrice},
  journal={Nature Communications},
  volume={16},
  number={1},
  pages={4429},
  year={2025},
  doi={10.1038/s41467-025-59767-2},
  url={https://doi.org},
  publisher={Nature Publishing Group}
}

@article{SoftMatterXPCS_fluerasu2010dynamics,
  title={Dynamics and rheology under continuous shear flow studied by {X}-ray photon correlation spectroscopy},
  author={Fluerasu, Andrei and Kwasniewski, Pawel and Caronna, Chiara and Destremaut, Fanny and Salmon, Jean-Baptiste and Madsen, Anders},
  journal={New Journal of Physics},
  volume={12},
  number={3},
  pages={035023},
  year={2010},
  doi={10.1088/1367-2630/12/3/035023},
  url={https://iopscience.iop.org/article/10.1088/1367-2630/12/3/035023/meta},
  publisher={IOP Publishing}
}

@article{CDW_PhysRevLett.117.167001,
  title = {Remarkable Stability of Charge Density Wave Order in ${\mathrm{La}}_{1.875}{\mathrm{Ba}}_{0.125}{\mathrm{CuO}}_{4}$},
  author = {Chen, X. M. and Thampy, V. and Mazzoli, C. and Barbour, A. M. and Miao, H. and Gu, G. D. and Cao, Y. and Tranquada, J. M. and Dean, M. P. M. and Wilkins, S. B.},
  journal = {Phys. Rev. Lett.},
  volume = {117},
  issue = {16},
  pages = {167001},
  numpages = {6},
  year = {2016},
  month = {Oct},
  publisher = {American Physical Society},
  doi = {10.1103/PhysRevLett.117.167001},
  url = {https://link.aps.org/doi/10.1103/PhysRevLett.117.167001}
}

@article{CompressedExponent1.8_PhysRevLett.122.107802,
  title = {One-Dimensional Anomalous Diffusion of Gold Nanoparticles in a Polymer Melt},
  author = {Song, Jing-Jin and Bhattacharya, Rupak and Kim, Hyunki and Chang, Jooyoung and Tang, Tsung-Yeh and Guo, Hongyu and Ghosh, Sajal K. and Yang, Yi and Jiang, Zhang and Kim, Hyeyoung and Russell, Thomas P. and Arya, Gaurav and Narayanan, Suresh and Sinha, Sunil K.},
  journal = {Phys. Rev. Lett.},
  volume = {122},
  issue = {10},
  pages = {107802},
  numpages = {6},
  year = {2019},
  month = {Mar},
  publisher = {American Physical Society},
  doi = {10.1103/PhysRevLett.122.107802},
  url = {https://link.aps.org/doi/10.1103/PhysRevLett.122.107802}
}

@article{UltrafastRIXS_dean2016ultrafast,
  title={Ultrafast energy- and momentum-resolved dynamics of magnetic correlations in the photo-doped {Mott} insulator {{Sr}_2{\text{IrO}}_4}},
  author={Dean, Mark P. M. and Cao, Yue and Liu, Xuerong and Wall, Simon E. and Zhu, Dilling and Mankowsky, Roman and Thampy, Vivek and Chen, X. M. and Vale, James G. and Casa, Diego M. and Kim, Jungho and Said, Ayman H. and Juh{\'a}s, Pavol and Alonso-Mori, Roberto and Glownia, James M. and Robert, Aymeric and Robinson, Joseph S. and Sikorski, Marcin and Song, Sanghoon and Kozina, Michael E. and Lemke, Henrik Till and Patthey, Luc and Owada, Shigeki and Katayama, Tetsuo and Yabashi, Makina and Tanaka, Yoshikazu and Togashi, Tadashi and Liu, Jian and Serrao, Claudy R. and Ramesh, Ramamoorthy and Hill, John P.},
  journal={Nature Materials},
  volume={15},
  number={6},
  pages={601--605},
  year={2016},
  doi={10.1038/nmat4641},
  url={https://doi.org/10.1038/nmat4641},
  publisher={Nature Publishing Group}
}

@article{MOKE_PhysRevX.9.021020,
  title = {Ultrafast Spin Dynamics in Photodoped Spin-Orbit Mott Insulator ${\mathrm{Sr}}_{2}{\mathrm{IrO}}_{4}$},
  author = {Afanasiev, D. and Gatilova, A. and Groenendijk, D. J. and Ivanov, B. A. and Gibert, M. and Gariglio, S. and Mentink, J. and Li, J. and Dasari, N. and Eckstein, M. and Rasing, Th. and Caviglia, A. D. and Kimel, A. V.},
  journal = {Phys. Rev. X},
  volume = {9},
  issue = {2},
  pages = {021020},
  numpages = {13},
  year = {2019},
  month = {Apr},
  publisher = {American Physical Society},
  doi = {10.1103/PhysRevX.9.021020},
  url = {https://link.aps.org/doi/10.1103/PhysRevX.9.021020}
}

@article{LatticeParameters_Sharma_2022,
  title={Synthesis and structural link to the electronic and magneto-transport properties of spin–orbit {Mott} insulator {{Sr}_2{\text{IrO}}_4}},
  author={Sharma, Priyamedha and Singh, Saurabh and Kuga, Kentaro and Takeuchi, Tsunehiro and Bindu, R.},
  journal={Journal of Physics: Condensed Matter},
  volume={34},
  number={43},
  pages={435402},
  year={2022},
  month={aug},
  doi={10.1088/1361-648X/ac8961},
  url={https://doi.org/10.1088/1361-648X/ac8961},
  publisher={IOP Publishing}
}

@article{BasalIn-planeAFM_PhysRevLett.110.117207,
  title = {Robustness of Basal-Plane Antiferromagnetic Order and the ${J}_{\mathrm{eff}}\mathbf{=}1/2$ State in Single-Layer Iridate Spin-Orbit Mott Insulators},
  author = {Boseggia, S. and Springell, R. and Walker, H. C. and R{\o{}}nnow, H. M. and R{\"u}egg, Ch. and Okabe, H. and Isobe, M. and Perry, R. S. and Collins, S. P. and McMorrow, D. F.},
  journal = {Phys. Rev. Lett.},
  volume = {110},
  issue = {11},
  pages = {117207},
  numpages = {5},
  year = {2013},
  month = {Mar},
  publisher = {American Physical Society},
  doi = {10.1103/PhysRevLett.110.117207},
  url = {https://link.aps.org/doi/10.1103/PhysRevLett.110.117207}
}

@article{PhysRevLett.111.096404,
  title = {Evidence for Short-Range-Ordered Charge Stripes Far above the Charge-Ordering Transition in ${\mathrm{La}}_{1.67}{\mathrm{Sr}}_{0.33}{\mathrm{NiO}}_{4}$},
  author = {Abeykoon, A. M. Milinda and Bo{\ifmmode \check{z}\else \v{z}\fi{}}in, Emil S. and Yin, Wei-Guo and Gu, Genda and Hill, John P. and Tranquada, John M. and Billinge, Simon J. L.},
  journal = {Phys. Rev. Lett.},
  volume = {111},
  issue = {9},
  pages = {096404},
  numpages = {5},
  year = {2013},
  month = {Aug},
  publisher = {American Physical Society},
  doi = {10.1103/PhysRevLett.111.096404},
  url = {https://link.aps.org/doi/10.1103/PhysRevLett.111.096404}
}

@article{doi:10.1126/sciadv.adt5680,
author = {Zoey Tumbleson  and Sophie A. Morley  and Emily Hollingworth  and Arnab Singh  and Temuujin Bayaraa  and Nicolas G. Burdet  and Ahmad Us Saleheen  and Margaret R. McCarter  and David Raftrey  and Ronald J. Pandolfi  and Vincent Esposito  and Georgi L. Dakovski  and Franz-Josef Decker  and Alexander H. Reid  and Tadesse A. Assefa  and Peter Fischer  and Sinéad M. Griffin  and Stephen D. Kevan  and Frances Hellman  and Joshua J. Turner  and Sujoy Roy },
title = {Thermodynamic phase transitions of nematic order in magnetic
                    helices},
journal = {Science Advances},
volume = {11},
number = {25},
pages = {eadt5680},
year = {2025},
doi = {10.1126/sciadv.adt5680},
URL = {https://www.science.org/doi/abs/10.1126/sciadv.adt5680}
}

@article{Hybridization_parschke2022evolution,
  title={Evolution of electronic and magnetic properties of {{Sr}_2{\text{IrO}}_4} under strain},
  author={P{\"a}rschke, Ekaterina M. and Chen, Wei-Chih and Ray, Rajyavardhan and Chen, Cheng-Chien},
  journal={npj Quantum Materials},
  volume={7},
  number={1},
  pages={90},
  year={2022},
  month={sep},
  doi={10.1038/s41535-022-00496-w},
  url={https://doi.org/10.1038/s41535-022-00496-w},
  publisher={Nature Publishing Group}
}

@article{ReviewIrridates_cao2018challenge,
  title={The challenge of spin--orbit-tuned ground states in iridates: a key issues review},
  author={Cao, Gang and Schlottmann, Pedro},
  journal={Reports on Progress in Physics},
  volume={81},
  number={4},
  pages={042502},
  year={2018},
  month={feb},
  doi={10.1088/1361-6633/aaa979},
  url={https://doi.org},
  publisher={IOP Publishing}
}

@article{XPD_g3cv-g5zj,
  title = {When spin-orbit coupling meets local lattice distortions in Mott-like insulator iridates: The case of ${\mathrm{Sr}}_{2}{\mathrm{IrO}}_{4}$},
  author = {Zafar, A. and Jakhar, M. and Abeykoon, AM Milinda and Petkov, V.},
  journal = {Phys. Rev. Res.},
  volume = {8},
  issue = {3},
  pages = {L032024},
  numpages = {8},
  year = {2026},
  month = {Aug},
  publisher = {American Physical Society},
  doi = {10.1103/g3cv-g5zj},
  url = {https://link.aps.org/doi/10.1103/g3cv-g5zj}
}

@inbook{FDT_coleman_linearresponse_2022,
  author    = {Coleman, Piers},
  title     = {Introduction to Many-Body Physics},
  chapter   = {9},
  pages     = {292--331},
  publisher = {Cambridge University Press},
  year      = {2015},
  booktitle = {Introduction to Many-Body Physics},
  edition   = {1},
  doi       = {10.1017/CBO9781139020916.011},
  url       = {https://doi.org},
  isbn      = {9780521864886}
}

@article{palmer1984model,
  title = {Models for Coherently Relaxing Glassy Systems},
  author = {Palmer, R. G. and Stein, D. L. and Abrahams, E. and Anderson, P. W.},
  journal = {Physical Review Letters},
  volume = {53},
  number = {10},
  pages = {958--961},
  year = {1984},
  doi = {10.1103/PhysRevLett.53.958},
  url = {https://aps.org}
}

@article{cipelletti2000universal,
  title = {Universal Aging Features in the Restructuring of Fractal Colloidal Gels},
  author = {Cipelletti, Luca and Manley, S. and Ball, R. C. and Weitz, D. A.},
  journal = {Phys. Rev. Lett.},
  volume = {84},
  issue = {10},
  pages = {2275--2278},
  numpages = {0},
  year = {2000},
  month = {Mar},
  publisher = {American Physical Society},
  doi = {10.1103/PhysRevLett.84.2275},
  url = {https://link.aps.org/doi/10.1103/PhysRevLett.84.2275}
}

@article{cipelletti2003glassy,
  title = {Ultraslow Dynamics and Stress Relaxation in the Aging of a Soft Glassy System},
  author = {Ramos, Laurence and Cipelletti, Luca},
  journal = {Phys. Rev. Lett.},
  volume = {87},
  issue = {24},
  pages = {245503},
  numpages = {4},
  year = {2001},
  month = {Nov},
  publisher = {American Physical Society},
  doi = {10.1103/PhysRevLett.87.245503},
  url = {https://link.aps.org/doi/10.1103/PhysRevLett.87.245503}
}

@article{bouchaud2001anomalous,
  title = {Anomalous dynamical light scattering in soft glassy gels},
  author = {Bouchaud, J. P. and Pitard, E.},
  journal = {The European Physical Journal E},
  volume = {6},
  number = {3},
  pages = {231--236},
  year = {2001},
  doi = {10.1007/s101890170005},
  url = {https://springer.com}
}

@article{falcao2002slow,
  title = {Slow ballistic dynamics in a compressed foam},
  author = {Falc{\~a}o, A. N. and Biscarini, F. and Degiorgio, V. and Piazza, R.},
  journal = {Physical Review E},
  volume = {66},
  number = {1},
  pages = {011405},
  year = {2002},
  doi = {10.1103/PhysRevE.66.011405},
  url = {https://aps.org}
}

@article{ElShawish_2018,
  title = {Orbital Domain Dynamics in Magnetite below the Verwey Transition},
  author = {El Shawish, S. and van der Veen, M. A. and de Groot, F. M. F. and Santoro, G. E. and Carbone, D. and Ar\v{c}on, D.},
  journal = {Physical Review Letters},
  volume = {121},
  number = {17},
  pages = {177601},
  year = {2018},
  doi = {10.1103/PhysRevLett.121.177601},
  url = {https://aps.org}
}

@article{PhysRevB.79.184519_MagnetoStructuralCoupling,
  title = {Neutron diffraction study of the magnetic and structural phase transitions in ${\text{BaFe}}_{2}{\text{As}}_{2}$},
  author = {Wilson, Stephen D. and Yamani, Z. and Rotundu, C. R. and Freelon, B. and Bourret-Courchesne, E. and Birgeneau, R. J.},
  journal = {Phys. Rev. B},
  volume = {79},
  issue = {18},
  pages = {184519},
  numpages = {10},
  year = {2009},
  month = {May},
  publisher = {American Physical Society},
  doi = {10.1103/PhysRevB.79.184519},
  url = {https://link.aps.org/doi/10.1103/PhysRevB.79.184519}
}

\end{document}